\documentclass{emulateapj}
\usepackage{apjfonts}
\usepackage{natbib}
\shorttitle{Diffuse hard X-ray emission in M82}
\shortauthors{Strickland \& Heckman}

\newcommand{\eg}{{\rm e.g.\ }}

\newcommand{\ie}{{\rm i.e.\ }}

\newcommand{\etal}{{\rm et al.\thinspace}}

\newcommand{\cm}{{\rm\thinspace cm}}
\newcommand{\km}{{\rm\thinspace km}}

\newcommand{\pcc}{\hbox{$\cm^{-3}\,$}}

\newcommand{\s}{{\rm\thinspace s}}
\newcommand{\yr}{{\rm\thinspace yr}}
\newcommand{\erg}{{\rm\thinspace erg}}
\newcommand{\ps}{\hbox{\s$^{-1}\,$}}
\newcommand{\pyr}{\hbox{\yr$^{-1}$}}
\newcommand{\ergps}{\hbox{$\erg\s^{-1}\,$}}

\newcommand{\kmps}{\hbox{$\km\s^{-1}\,$}}
\newcommand{\pcmsq}{\hbox{$\cm^{-2}\,$} }

\newcommand{\halpha}{H$\alpha$}

\newcommand{\K}{{\rm K}}
\newcommand{\hi}{H{\sc i}}

\newcommand{\nH}{\hbox{$N_{\rm H}$}}
\newcommand{\Mdot}{\hbox{$\dot M$}}
\newcommand{\Edot}{\hbox{$\dot E$}}

\newcommand{\kpc}{{\rm\thinspace kpc}}
\newcommand{\Mpc}{{\rm\thinspace Mpc}}
\newcommand{\keV}{{\rm\thinspace keV}}

\newcommand{\Lsol}{\hbox{$\thinspace L_{\sun}$}}
\newcommand{\Msol}{\hbox{$\thinspace M_{\sun}$}}

\begin{document}

\title{Iron line
and diffuse hard X-ray emission from the starburst galaxy M82}

\author{David K. Strickland \altaffilmark{1} and
        Timothy M. Heckman.\altaffilmark{1}}

\altaffiltext{1}{Department of Physics and Astronomy,
        The Johns Hopkins University,
        3400 N.~Charles St., Baltimore, MD 21218, USA.}


\begin{abstract}
We examine the properties of the diffuse hard X-ray emission in the 
classic starburst galaxy M82.
We use new {\it Chandra} ACIS-S observations in combination with reprocessed
archival {\it Chandra} ACIS-I and {\it XMM-Newton} 
observations. We find $E \sim 6.7$ 
keV Fe He$\alpha$ emission (from highly ionized iron) is present
in the central $|r| < 200$ pc, $|z| < 100$ pc of 
M82 in all datasets at high statistical
significance, in addition to a possibly non-thermal X-ray continuum
and marginally significant $E=6.4$ keV Fe $K\alpha$ line
emission (from weakly ionized iron). 
No statistically significant
Fe emission is found in the summed X-ray spectra of the point-like X-ray
sources or the ULX in the two epochs of {\it Chandra} observation.
The total nuclear region iron line fluxes in the 2004 April 21 
XMM-Newton observation
are consistent with those of the Chandra-derived diffuse component,
but in the 2001 May 6 XMM-Newton observation they are significantly higher 
and also both E=6.4 and E=6.9 keV iron lines are detected. We attribute
the excess iron line emission to the Ultra-Luminous X-ray source
in its high state. In general the iron K-shell luminosity of M82 is
dominated by the diffuse component.
The total X-ray luminosity of 
the diffuse hard X-ray emission (corrected for emission by unresolved
low luminosity compact objects) is
$L_{\rm X,2-8 keV} \sim 4.4\times10^{39} \ergps$ in the $E=2$ -- 
8 keV energy band, and the 6.7 keV iron line luminosity is
 $L_{\rm X, 6.7 keV} \sim $(1.1 -- 1.7) $\times 10^{38} \ergps$.
The diffuse hard X-ray continuum is best fit with power law models
of photon index $\Gamma \sim 2.5$ -- 3.0, although thermal bremsstrahlung 
model with $kT \sim 3$ -- 4 keV are also acceptable. 
A simple interpretation of the hard diffuse continuum is
that it is the bremsstrahlung associated with the 6.7 keV iron line
emission. This interpretation is problematic as it would imply
an unrealistically low gas-phase iron abundance 
$Z_{\rm Fe} \sim 0.4 Z_{\rm Fe, \odot}$, much lower than the
super-Solar
abundance expected for supernova ejecta. We explore the possibility
of non-equilibrium ionization, but find that this can not
reconcile the continuum and iron line emission.
Nor can Inverse Compton X-ray emission be the dominant source
of the diffuse continuum as theory predicts flatter power law slopes
and an order of magnitude less continuum flux than is observed.
The 6.7 keV iron line luminosity is consistent with that expected from
the previously unobserved 
metal-enriched merged supernova ejecta that is thought to drive the
larger-scale galactic superwind. The iron line luminosity 
implies a thermal pressure
within the starburst region of $P/k \sim 2 \times 10^{7} \K \pcc$
which is consistent with independent observational estimates of
the starburst region pressure.
\end{abstract}

\keywords{ISM: jets and outflows --- ISM: bubbles ---
galaxies: individual : NGC 3034 (M82) 
--- galaxies: halos --- galaxies: starburst --- X-rays: galaxies}



\section{Introduction}
\label{sec:introduction}

Starburst galaxies -- objects with intense recent star formation --
have galaxy-sized outflows
with velocities in excess of several hundred to a thousand
kilometers per second. These superwinds
are common in the local Universe, occurring in nearly all
galaxies classified as undergoing starburst activity
\citep{lehnert95,heckman98},
and appear ubiquitous among the Lyman break
galaxies at redshift $\sim 3$ where they are blowing $\sim 100$ kpc
sized holes in the IGM \citep[\eg][]{pettini01,adelberger03}.

Superwinds are believed to be driven by the thermal
and ram pressure of an initially very hot ($T\sim 10^{8}$ K), 
high pressure ($P/k ~ \sim 10^{7} \K \pcc$) and
low density wind, itself created from the merged
remnants of very large numbers of core-collapse supernovae (SNe),
and to a lesser
extent the stellar winds from the massive stars, that occur
over the $\sim 100$ Myr duration of a typical starburst event
\citep{chevclegg}. Direct observational detection of gas this
hot through X-ray spectroscopy or 
imaging has only recently become possible because of the
high spatial resolution of {\it Chandra} X-ray Observatory, 
as starburst regions 
are also home to large numbers
of X-ray-luminous accreting neutron stars and black holes that must
be resolved out in order to see any diffuse hard X-ray emission.


All known superwinds were detected through observations of ambient
disk or halo gas that has been swept-up by the hotter
but more tenuous wind of merged SN ejecta\footnote{It is useful draw a 
conceptual distinction between
the material that actually drive superwinds, 
and the many other phases of swept-up
and entrained gas that are part of superwinds. We shall define the wind-fluid
as the merged SN ejecta (possibly mass-loaded by ambient gas) whose
high thermal and/or ram pressure drives the superwind.}, which has itself
escaped direct observational detection until recently. 
The majority of nearby superwinds have been discovered
using optical imaging and spectroscopy to identify outflow in the 
warm ionized medium  (WIM, $T\sim10^{4}$ K), 
in many cases directly imaging bipolar structures
aligned with the host galaxy minor axis that have kinematics indicative
of outflow at velocities of $v_{\rm WIM} = $ a few $\times 100$ to $1000 \kmps$
\citep*[\eg][]{axon78,ham87,bland88,ham90,lehnert95,lehnert96a}.
\citet{mccarthy87}, \citet{ham90}, and \citet{lehnert96a} 
use this entrained low-filling factor gas as a tracer of the
pressure in the inner kiloparsec or so of many nearest brightest superwinds,
and find the very high central pressures and radial variation in pressure
expected from the \citet{chevclegg} wind model for the thermalized
SN ejecta within the vicinity of the starburst region.
Superwinds from galaxies at high redshift
are recognized by virtue of blue-shifted interstellar
absorption lines from warm-neutral and warm-ionized gas
species \citep{pettini00,fbb02}, 
absorption features very similar to those seen
in local starburst galaxies with superwinds 
\citep{phillips93,kunth98,gonzalezdelgado98,heckman2000}. Nearby superwinds
have also been extensively studied in soft X-ray 
emission\footnote{We will use the following definitions in this paper: 
Soft X-ray denotes photon energies in the range 
$E \sim 0.1$ -- 2.0 keV. Hard X-ray denotes photon energies $E > 2 \keV$,
and within the range $2 < E < 10 \keV$ unless it is stated otherwise.}
\citep[\eg][]{dwh98,strickland04a} which probes emission from
hot gas with temperatures in the range $T \sim10^{6}$ to $10^{7} \K$, 
following their initial detection by
{\it Einstein} X-ray observatory \citep*{watson84,fabbiano84}.
Indeed, observations have demonstrated
that all phases of the ISM found in normal late type galaxies are 
also incorporated into starburst-driven superwinds 
\citep[see the review of][]{dahlem97}.

However, prior to the launch of the {\it Chandra} X-ray
Observatory (CXO) in 1999, there had been
no believable detection of thermal X-ray 
emission from the $T\sim$
 a few $\times 10^{7}$ to $10^{8}$ K merged SN ejecta whose
thermal pressure initially drives the superwind 
in any starburst galaxy. 
This very hot plasma should be a source of thermal bremsstrahlung 
in the hard X-ray energy band, as well as strong 
$E \approx 6.7$ and (possibly) $6.9$ keV emission from 
highly-ionized Helium-like and Hydrogen-like iron.
Hard X-ray emission had been previously been 
detected from nearby starburst galaxies
such as M82 and NGC 253 with X-ray observatories
such as {\it EXOSAT} \citep{schaaf89}, {\it Ginga} \citep{tsuru90}, 
{\it ASCA} \citep{ptak97,tsuru97} and {\it BeppoSAX} \citep{persic98,cappi99}. 
However, the poor-to-nonexistent
spatial resolution of these observations prevented a clean assessment
of the relative contribution to the hard X-ray emission from
compact accreting objects and from any truly diffuse component 
(the half power enclosed angular diameter of the point spread function (PSF)
of {\it BeppoSAX} and {\it ASCA} 
at $E \sim 6 \keV$ were $r_{50} \sim 2\arcmin$ and
$\sim 3\arcmin$ respectively. This was the highest spatial resolution 
available for any hard X-ray observatory before the launch of {\it Chandra}.
Assuming a distance of 2.6 Mpc to NGC 253 and
3.6 Mpc to M82, an angular size of $2\arcmin$ corresponds to a physical
size of $1.5\kpc$ for NGC 253 and $2.1 \kpc$ for M82).

Nevertheless it was clear that the majority of the hard X-ray
emission from these galaxies came from compact objects. The
observed total hard X-ray luminosities of these galaxies were
$\ga 1.5$ orders of magnitude higher than expected from the
the very hot gas in a superwind model \citep{ss2000}.
Both the {\it ASCA} and {\it BeppoSAX} observations of M82 and NGC 253 
also discovered emission possibly 
due to high ionized iron \citep{ptak97,tsuru97,persic98,cappi99}, 
although this emission was of low equivalent width.
The weakness of the Fe K-shell emission implied 
sub-Solar Fe abundances ($Z_{Fe} \la 0.3 Z_{Fe,\odot}$), 
a peculiar result given that 
super-Solar abundances ($Z_{Fe} \sim 5 Z_{Fe,\odot}$)
are expected from pure core-collapse SN ejecta. Alternatively
the hard X-ray emission could be dominated by 
non-thermal continuum emission, possibly from X-ray binaries, thus 
reducing the equivalent width of the Fe line from the diffuse hot gas. 
Furthermore,
some Galactic X-ray binaries do show Fe K-shell emission 
\citep*[\eg see][]{vsr93}, so
the Fe emission observed in M82 and NGC 253 might come from compact
objects rather than any diffuse gas. 
Despite the low spatial resolution of {\it ASCA}, the hard X-ray emission
in NGC 253 was distributed so broadly over the disk of the galaxy that
\citet{ptak97} were able to resolve multiple individual hard X-ray
sources. These were also directly associated with point-like X-ray
sources seen in soft X-ray observations with the {\it ROSAT} High Resolution
Imager (HRI). The strong temporal variability in 
the hard X-ray luminosity of M82 \citep{ptak99,matsumoto99}
demonstrated that one or more accreting compact objects
dominated the hard X-ray emission at some, and possibly all, epochs.

 The on-axis half-enclosed-power PSF diameter
of Advanced CCD Imaging Spectrometer (ACIS) on the 
{\it Chandra X-ray Observatory}  is $0\farcs8$, a factor of $150$ smaller than BeppoSAX,
and also considerably better than the equivalent value of $\sim 13$ -- 
$15\arcsec$ for the MOS and PN detectors on {\it XMM-Newton}.
The angular diameter of the starburst region in M82 is 
$\sim 45\arcsec$ \citep*[see \eg][]{gak96}, while that of NGC 253
is $\la 15\arcsec$ \citep{ua97,forbes00}.
It is  clear that {\it Chandra} is the first, and only, X-ray
instrument with spatial resolution high enough to resolve
the starburst regions in even the nearest starburst galaxies
such as M82 and NGC 253.

Using the spectral-imaging CCDs
of the ACIS-I camera of {\it Chandra}, \citep{griffiths2000}
detected \emph{diffuse} hard X-ray emission 
coincident with the nuclear starburst region of M82. 
They interpreted the spectrum of this
diffuse emission at thermal bremsstrahlung from a hot plasma
of temperature $T \sim 4\times10^{7} \K$. This was motivated by the
detection of excess emission in the energy range
$E = 5.9$ to 6.9 keV which they attributed to unresolved line emission from
highly-ionized iron. This feature that would not be expected if
the diffuse X-ray emission where non-thermal in origin, for
example inverse Compton (IC) scattering of IR photons off relativistic
electrons. This work was based on ACIS-I 
observations taken in late 1999, soon after the cosmic-ray-induced 
damage to front-illuminated ACIS-I CCDs had been discovered.
As a result, the spectral calibration of these 
observations was especially difficult. In particular, the
radiation-induced reduction in spectral resolution and absolute
energy scale calibration of the instrument made it difficult to be sure that
Fe emission feature was indeed real.

ACIS-S observations also uncovered diffuse hard X-ray 
emission in the compact nuclear starburst region of NGC 253 \citep{weaver02}.
The emission is strongest within the central $5\arcsec$ of the
galaxy, and showed a strong hard X-ray emission line spectrum with features
attributable to Si, S, Ca and Fe (the back-illuminated 
chips on ACIS-S suffered
less radiation damage than the front-illuminated chips, and hence
have higher spectral resolution). This spectrum is more reminiscent
of a highly photoionized plasma than a collisionally ionized plasma. 
As there is independent evidence for a low luminosity AGN (LLAGN)
in the center of NGC 253 \citep{ua97}, \citet{weaver02} suggest
that the diffuse hard X-ray emission from NGC 253 is dominated by
gas photoionized by the LLAGN.

Thus the nature of this diffuse hard X-ray emission in M82, and
its possible relationship to the very hot merged SN ejecta thought
to drive M82's superwind, has not yet been resolved and is the focus
of this paper.
We emphasize that M82 is the best case for exploring 
the properties of diffuse very hot gas created by starburst activity, 
as the diffuse hard X-ray emission is brighter and has
a larger angular extent than in NGC 253. Furthermore, there is
no known LLAGN in M82 to complicate matters.

In this paper we make use of the longest available {\it Chandra} ACIS
and {\it XMM-Newton} observations and the most recent
instrumental calibrations to further investigate the properties
of the diffuse hard X-ray emission in this archetypal starburst galaxy. 
With {\it Chandra} we spatially separate
 the diffuse hard X-ray emission 
from resolved point source emission, and use the {\it XMM-Newton} observations
provide the best signal-to-noise spectra of the nuclear starburst region
of the galaxy. These observations also sample
four different epochs, allowing us to separate the time-constant diffuse
component of the hard X-ray iron line emission from a one-time
contribution by the Ultra-Luminous X-ray source found in the shorter
{\it XMM-Newton} observation.

\section{Observations and data reduction}
\label{sec:data_analysis}

We obtained a 18 ks-long observation of M82 on 2002 June 18 (1061 days
since the launch of {\it Chandra}), with
the central few kpc of the galaxy on the back-illuminated S3
chips of the {\it Chandra} ACIS-S instrument 
(observation ObsID 2933). This ACIS-S observation
has  higher spectral resolution than the ACIS-I observations
presented in \citet{griffiths2000}.
In addition this new observation,
we have reanalyzed two archival ACIS-I observations of M82,
both taken on 1999 September 20 (55 days since launch, ObsIDs 1302 and
361, PI Murray and Garmire respectively),
  making use of the latest spectral
calibrations, and corrections for the radiation-damage induced
charge transfer inefficiency (CTI) in the ACIS-I chips
that were unavailable at the time of \citeauthor{griffiths2000} study.
The longer of these two ACIS-I observations (ObsID 361, approximately
33 ks in length) was analyzed by \citet{griffiths2000}.

In addition to reanalyzing this dataset, we have also combined 
the ObsID 1302 and ObsID 361 data into a single longer
dataset. Although image-based analysis of multiple merged {\it Chandra}
datasets is common in the literature, it is normally inadvisable to perform
spectral analysis on such merged data. In this particular case this
is not a concern as the observations were performed concurrently,
and the pointing directions and telescope roll of the observations are
essentially identical. As photons within the nuclear region of interest
fell on the same pixels of the same chips in both observations, the
chip-location-specific instrument responses are the same and we
can proceed with a spectral analysis on the merged data.

For convenience we shall refer to the different
datasets analyzed here as the ACIS-S (obsID 2933), 33 ks 
ACIS-I (ObsID 361) and merged ACIS-I (ObsID 361 + ObsID 1302) 
observations.

The Chandra data was reduced and analyzed in largely 
the same ways as described
in \citet{strickland04a}, although using more recent
versions of the software and improved spectral and
energy calibrations.
A few details of the data reduction of
specific relevance to this paper are discussed here,
in particular the differences from the the \citet{strickland04a}
data reduction. 

We also reduced and analyzed the data for the two longest
{\it XMM-Newton} observations of M82, obtaining the
data from the public archive. The EPIC instruments
on {\it XMM-Newton} do not have high enough
spatial resolution to separate the diffuse and point source related
X-ray emission in the nucleus of M82. We make use of EPIC's higher
effective collecting area
to place high signal-to-noise constraints on the \emph{total} 
K-shell iron line emission from M82 at different
epochs from the {\it Chandra} observations. 

The {\it Chandra} and {\it XMM-Newton} observations are
summarized in Table~\ref{tab:obslog}.

We adopt a distance of $3.6 \Mpc$ to M82. This is based
on the Cepheid distance of $3.63\pm{0.34} \Mpc$
to its close neighbor galaxy M81 
\citep{freedman94}, to which M82 is connected by \hi~tidal tails
\citep{cottrell77}. This distance is consistent with detection of
the tip of the red giant branch in M82, which indicates
a distance of $3.9\pm{0.4}\Mpc$ \citep{sakai99}. At our adopted
distance an angular size of $1\arcsec$ corresponds to a physical
size of 17.5 pc. We also adopt a position angle of $73\degr$
for the major axis of the galaxy \citep{achtermann95} and the
location of the dynamical center of the galaxy at 
$\alpha=09^{\rm h} 55^{\rm m} 51\fs9, 
\delta=+69\degr 40\arcmin 47\farcs1$ (J2000.0) 
from \citet{weliachew84}.

\begin{deluxetable*}{lllrll}
\tablecolumns{6}
\tablewidth{0pc}
\tablecaption{{\it Chandra} and {XMM-Newton} observations of M82
        \label{tab:obslog}}
\tablehead{
\colhead{Date} & \colhead{Observatory} 
  & \colhead{Instrument} & \colhead{ObsID} & \colhead{Exposure time} 
  & \colhead{Name used} 
}
\startdata
1999-09-20 & {\it Chandra}    & ACIS-I & 361        & 15388 (I3)
  & (not used) \\
1999-09-20 & {\it Chandra}    & ACIS-I & 1302       & 32932 (I3) 
  & 33ks ACIS-I \\
1999-09-20 & {\it Chandra}    & ACIS-I & 361+1302   & 48320 (I3) 
  & merged ACIS-I\\
2001-05-06 & {\it XMM-Newton} & EPIC   & 0112290201 & 23327 (PN), 
  29685 (MOS1), 29687 (MOS2)
  & XMM-short \\
2002-06-18 & {\it Chandra}    & ACIS-S & 361        & 17912 (S3) 
  & ACIS-S \\
2004-04-21 & {\it XMM-Newton} & EPIC   & 0206080101 & 59990 (PN), 
  73637 (MOS1), 73629 (MOS2)
  & XMM-long \\
\enddata
\tablecomments{The ObsID is the unique observation identification number
  assigned by the {\it Chandra} and {\it XMM-Newton} Science Centers.
  The exposure times (in seconds) quoted are those remaining after
  the removal of periods of higher-than-normal background. The specific chip
  (for {\it Chandra} observations) or detector (for {\it XMM-Newton}) the
  exposure time applies to is noted in parentheses.
}
\end{deluxetable*}

\subsection{Chandra data}

\subsubsection{CTI and gain correction}

The data was reduced using CIAO
(version 3.2). The latest calibrations from
version 3.0 of the Chandra calibration database (CALDB).
Note that this version of CIAO can correct for the radiation-induced
CTI (manifest as an apparent decrease photon event energy, 
and spectral resolution, with increasing
distance from the CCD readout) 
of the front-illuminated ACIS CCDs, but not the back-illuminated
CCDs, for observations
with the focal plane instruments at $-120\degr$ C. However, the
effect of CTI in the ACIS-S
observations of M82 is mitigated by the placement of the nuclear region
within $\sim1/3$ of a chip-width from 
the readout edge of the S3 chip, and the relative lack
of radiation damage to the back-illuminated chips S1 and S3.
The changes in apparent photon energy due to ongoing increases in CTI in 
the the back-illuminated CCDs is largely corrected for in CIAO 3.2
by the incorporation of time-dependent gain corrections\footnote{Vikhlinin 
\etal, 2003, Corrections for time-dependence of ACIS gain, 
\url{http://hea-www.harvard.edu/~alexey/acis/tgain/}}.

CIAO 3.2 can
 not CTI-correct the earlier ACIS-I observations of M82, which were taken
with a focal plane temperature of $-110\degr$ C. In these observations the
nucleus of M82 is at the far edge of the I3 chip from the readout edge,
maximizing the negative effects of CTI. We used the CTI-correction
software (version 1.45) developed by 
the Penn State University (PSU) ACIS group \citep[see][]{townsley00}
to CTI-correct
the M82 ACIS-I data, before reducing the data using CIAO as normal.
This corrects for the position-dependent energy degradation, including the
secular increase in CTI discussed above, but does not
correct for decrease in spectral energy resolution also associated with
CCD CTI. For this reason the CTI-corrected
ACIS-I observations of M82 have lower spectral
resolution than the ACIS-S observations.

The standard {\it Chandra} data pipe-line introduces an artificial
0\farcs5 Gaussian randomization of photon positions. We remove this
in our reduction of the data. In \citet{strickland04a} we applied
a technique called PHA randomization to remove an artificial
quantization in the raw photon pulse height amplitude (PHA) values. CTI
and time-dependent gain corrections automatically remove this
quantization, making CIAO PHA randomization unnecessary (L.~Townsley, 
private communication). Removal of events at node-boundaries and standard event
grade filtering was performed to further reduce the number of
non-X-ray events in the data.

Following gain or CTI correction, both the ACIS-S and ACIS-I datasets were
processed to remove artificial 
streaks of events along the same CCD columns as the brightest point source
in M82. This streak is caused by photons from bright point sources that
are detected during the $40 \mu$s
time period while the chip is being read out after each 3.2 s exposure.
Corrections were applied to each ACIS data set to remove the event streaks
expected from the six brightest point sources within the nuclear region
of M82.

In order to achieve the most-accurate absolute astrometric solution for
these observations the latest satellite aspect solutions were applied
to the data\footnote{See \url{http://asc.harvard.edu/ciao/threads/arcsec\_correction/}.}. 
With these sub-arcsecond corrections applied, the astrometry
of the data should be accurate to
$0\farcs6$ at 90\% confidence.

\subsubsection{Flare-removal and source searching}

We used the techniques for removing periods of higher-than-normal
particle background (flares), and point source detection and removal,
described in \citet{strickland04a}. We refer the reader
to that work for a detailed description.

After flare removal, the exposure time on the S3 chip in the ACIS-S 
observation of M82 is $t_{exp}=17.912$ ks, $t_{exp}=32.932$ ks
on the I3 chip of the 33ks ACIS-I observation and $t_{exp}=48.320$ ks
on the I3 chip of the merged ACIS-I observation.

\subsubsection{Background subtraction}
\label{sec:data_analysis:background}

The diffuse hard X-ray emission from the central region of M82
covers a region of sky $\la 1\arcmin$ 
in angular diameter. In this aperture the background (a 
sum of the extragalactic and galactic X-ray background emission and 
particle events in the detector) is weak but non-negligible.


Soft diffuse X-ray emission from the superwind covers the majority
of the S3 chip in the ACIS-S observation 
(see Fig.~3 in \citealt{strickland04a}), leaving little area ($\la 5.4$
arcmin$^{2}$) from
which to extract a high signal to noise background spectrum.
We instead used the blank-sky datasets 
produced by Maxim Markevitch\footnote{Markevitch. 2001, 
General discussion of the quiescent and flare components of 
the ACIS background, \url{http://cxc.harvard.edu/contrib/maxim/bg/index.html}.
We used the 2004 December 12 version of the ACIS background datasets.}
for background subtraction.
These include CTI and gain correction
consistent with our ACIS-S observation of M82, 
unlike the blank sky datasets that are part of CALDB 3.0.

We used data from regions on the other ACIS chips that are free of
point source or diffuse X-ray emission in our observation 
(specifically chips S1, S2 and S4) to (a)
check that the spectral shape of the background in our observation is
consistent with that in the blank-sky datasets, and (b)
calculate the factor by which the blank sky background data needs to be
scaled to match the background in our observation. The back-illuminated
S1 chip is the most useful chip in this regard. In the ACIS-S observation the
S1 chip is offset to the SSE of the nucleus of M82, along the minor axis
of the galaxy. The edge of the chip nearest to M82 is $\sim10\arcmin$
from the nucleus, well beyond the edge of any diffuse X-ray or other
emission associated with the superwind. Furthermore, its sensitivity to
background events is very similar to that of the S3 chip, unlike 
the front-illuminated S2, S4 and S5 chips of the ACIS-S. We found that
both the absolute normalization and 
spectral shape of the observed S1 background matched that in the
blank-sky background datasets very closely. The background
normalization in our M82 observation was marginally lower than
in the blank-sky datasets (by $1.9\pm{1.4}$\% in the 5 -- 8 keV energy band,
$1.2\pm{1.2}$\% in the 2 -- 8 keV band, using an area of $\sim 60$ 
arcmin$^{2}$. These values are consistent the $\sim 2$\% rms scatter in the
normalization of the hard X-ray component from observation to observation
\citep{hickox06}. Their measurements used a higher energy range 
than the one we have used, but as they also show that the spectrum
of the background very stable we conclude that a 1.9\% renormalization
is within the expected levels of variation.

At lower energies, or in 
narrower energy bands the constraints are weaker and hence consistent with
no difference in normalization). There is known soft diffuse X-ray emission
at the location of the S2 and S4 chips, but in the hard X-ray band the observed
and blank-sky background surface brightness levels in these chips
are also consistent
with each other.
The resulting blank-sky background dataset for the S3 chip had an
effective exposure of 458.60 ks, 
when the 1.9\% renormalization described above was been applied.

The resulting background dataset has the same pointing and orientation
as the actual observations of M82. Background subtraction is accomplished 
simply by subtracting from the observation data (be it an image or spectrum) 
the equivalent 
data from the background dataset in the same spatial region and energy
band, scaled by the ratio of exposure times. The background is typically
small compared to the true emission from M82 (see Table~\ref{tab:diff_frac}).
At energies between $E=5$ -- 7 keV the background within a radius of
$\sim 0\farcm5$ of the center of M82 
is approximately $\sim 1$\% of the total emission and $\sim 4$\% of the
diffuse emission.

This method ignores the fact that some of the extragalactic X-ray
background will be absorbed by M82's disk (\ie M82 will cast a shadow in the
X-ray background), resulting in a slight
over-subtraction of the background. \citet{bregman02} demonstrated such
shadowing of the soft X-ray background by the disk of the edge-on spiral
NGC 891. Shadowing is only visible at low energies ($E \la 1$ keV),
where the total {\it Chandra} ACIS-S background (instrumental particle
events as well as the Galactic and extragalactic X-ray backgrounds) is
reduced to $\sim 70\pm{10}$\% of the value found in regions not covered
by the galaxy. Thus we may have over-estimated the background in
some regions by $\sim 30$\%. Given that the background is so 
small compared to the
diffuse emission we are interested in this systematic error is
negligible compared to statistical uncertainties in the diffuse
flux (Table~\ref{tab:diff_frac}). Furthermore the 
shadowing of the extragalactic
background will be greatest toward the center of M82, but the relative 
significance of this systematic over-subtraction is mitigated by the
intrinsically higher emission from M82 in this region.


For the ACIS-I observations of M82 background subtraction is slightly more
straightforward. Moderately large fractions of each of the four ACIS-I
chips lie beyond the diffuse soft X-ray emission associated with the superwind,
and hence are used to for background subtraction. All of the hard diffuse
X-ray emission lies within the I3 chip, for which 
a background region of area $18$ square arcminutes free of
diffuse and point-source X-ray emission is available.

This local background estimate needs to be scaled appropriately when
calculating the expected background in any other region on the same ACIS-I 
chip, given that the background will change with position. 
For image-based analysis
we used the relative intensity from an exposure map (the product of the
effective exposure, and collecting area, at some X-ray photon energy for
each point on each ACIS chip)
to map from an observed local background to a chip-wide model background
image. This was done on a chip-by-chip and energy band-by-energy band basis.

For spectral analysis we do not use this exposure-map based correction.
We instead scale the spectrum from the background region by the ratio
of the area of the source  region (the center of M82) to the larger
background region. No correction for vignetting or shadowing of the
background is applied, but this does not introduce significant
systematic errors because the background is so small compared to the
diffuse emission.

\subsubsection{Spectral fitting}

Spectral fitting was performed using
XSPEC (version 11.3.1t). Spectral response files were created for
the ACIS-S data using CIAO and the latest CALDB. The instrument
effective area takes into account the 
time-and-spatially-dependent
decrease in low-energy sensitivity due to molecular contamination
of the optical blocking filter (OBF), as well as the weak spatial
dependence of the effective area.
For the ACIS-I observation we used the ACIS-I3 aim-point RMF file
provided as part of the PST CTI correction software, and created
the ARF file with the CIAO task {\sc mkarf} using the QEU file
provided with the PSU software.

For the ACIS-S observation a background spectrum was extracted from
the blank-sky background dataset in
a $1\arcmin$-diameter circle centered on the dynamical center of M82
(see \S~\ref{sec:results:spectra}). 
Note that using the blank-sky background datasets automatically takes
account of the spatial dependence of the background. For the ACIS-I 
observation the background spectrum was taken from the large 
emission-free region on the I3 chip mentioned above.

Background-subtracted spectra were manually created using the CIAO
task {\sc dmtcalc}, instead of letting the {\sc Xspec}
spectral-fitting program
perform the background subtraction itself.
This allows us to both re-bin the spectra to
ensure a minimum of 10 counts per bin after background subtraction
(using {\sc dmgroup}),
and associate more-realistic statistical errors with the data
(the \citet{gehrels86} approximation to Poissonian errors,
instead of $\surd N$ errors).

\subsubsection{The effect of new calibration products}

We investigated the effect these recent calibrations have
on our results, compared to those based on the CIAO 2.2, 
CALDB 2.12, non-CTI-corrected
data reduction used in \citet{strickland04a}. There is no significant
difference in conclusions based on image analysis using these
different calibrations. This is to be expected, as the photon energy range
in most images is much larger than the changes to photon energy
resulting from the
CTI or gain corrections. Where the new calibrations do make a 
significant difference is in spectral fitting, in particular to the
energy of line features or line-complexes. For example, we find the 
mean energy of the Fe-K feature in the ACIS-S observation
nuclear spectrum is $\sim 0.1 \keV$ higher when using the most
recent calibration data and gain corrections.


\begin{figure*}[!ht]
\epsscale{1.1}
\plotone{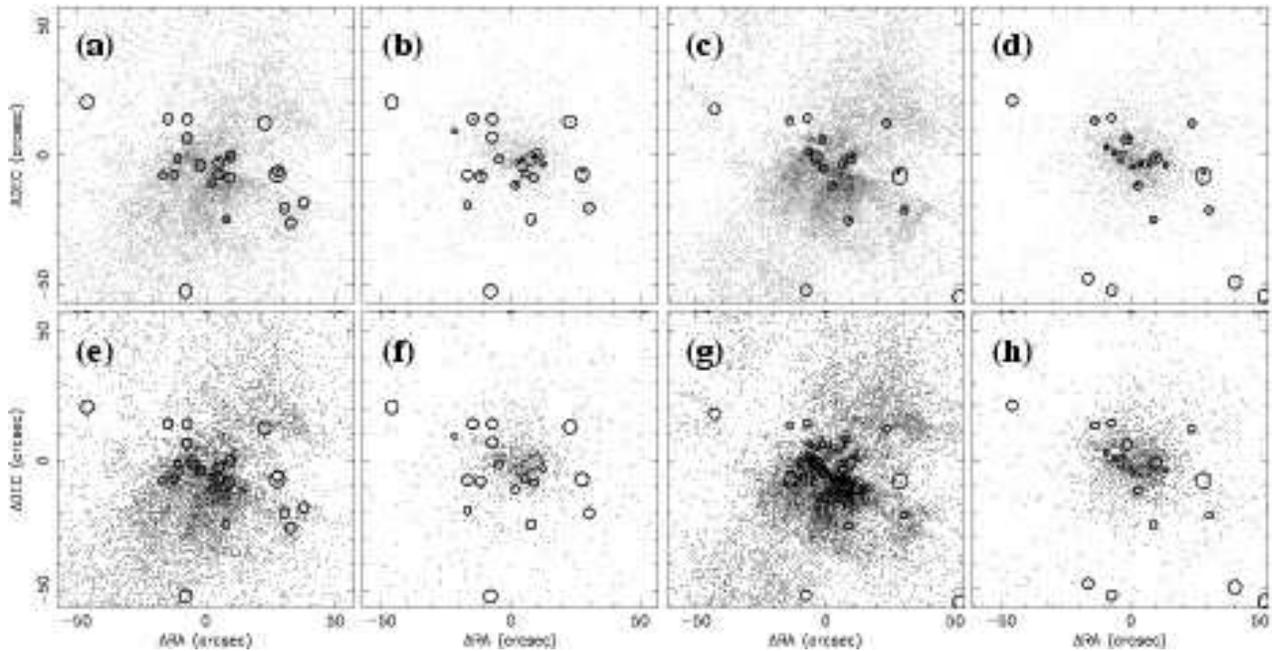}
  \caption{The center of
  M82 as seen with {\it Chandra}. Each image
 shows the raw, unsmoothed, photon distribution in the soft
  0.3--2.0 keV (panels a and c) and and hard 2.0--8.0 keV 
  energy bands (panels b and d), binned in $0\farcs492$-wide 
  pixels, over a $2\times2$ kpc ($1\farcm90 \times1\farcm90$) region
  centered on the dynamical center of the galaxy. 
  Panels a and b are from the ACIS-S observation, while panels
  c and d are from the merged ACIS-I observation. The images are
  shown on a square-root intensity scale running from 0 to 75 counts
  per pixel. Point-like X-ray sources detected in the soft and hard
  energy bands are shown surrounded by a circle, equivalent in size
  to the region used to remove the point sources from the images 
  and spectra.
  Panels e, f, g and h are equivalent to panels a, b, c and d respectively,
  except that point sources have been removed and interpolated over,
  and the intensity scale runs from 0 to 25 counts per pixel. The background
  has been subtracted from all the images in the manner described in
  \S~\ref{sec:data_analysis:background}.
  }
  \label{fig:xim_both_obs}
\end{figure*}


\subsection{XMM-Newton observations}

The two longest {\it XMM-Newton} observations of M82 
were performed on 2001 May 06 (ObsID 112290201, 
PI Martin Turner, initial duration 29.4 ks, 
henceforth referred to as ``XMM-short'') 
and on 2004 April 21 (ObsID 206080101, initial duration 105.4 ks,
PI Piero Ranalli, which we
shall refer to as ``XMM-long'').
We did not use a shorter observation ($\sim 10$ ks) taken on the
same date as the XMM-short observation because of the low
S/N of the resulting spectra.

We reprocessed the archived data using version 6.5 of the {\it XMM-Newton}
SAS software, and reran the EPIC MOS and PN analysis chains
with the latest calibration data. The recommended standard event grade 
filters and proton flare screening was applied to each dataset.
After filtering the remaining exposure times in the XMM-short observation
were 21.1 ks and 29.4 ks for PN and each MOS instruments respectively,
and in the XMM-long observation 54.0 ks and 72.8 ks for the 
PN and each MOS detectors respectively.

Spectra and response files were created using SAS using the methods
described in the on-line SAS Data Analysis 
Threads\footnote{See \url{http://xmm.vilspa.esa.es/sas/new/documentation/threads/}}. 

Nuclear-region spectra were extracted from the same area used for the
Chandra ACIS spectra (see \S~\ref{sec:results:spectra:chandra}).
Background spectra for each observation 
were accumulated in regions of radius $1\farcm25$ that were free of
point sources or diffuse X-ray emission. The same background
region was used for all three detectors (PN, MOS1 and MOS2) and 
whenever possible was on the same
CCD chip as the nuclear diffuse emission.

\section{Image analysis of the Chandra data}
\label{sec:results}

\subsection{The spatial distribution of the diffuse soft and hard X-ray emission}
\label{sec:results:images}

The spatial distribution of the soft and hard X-ray emission within 
$\sim1$ kpc of the center of M82 as seen in the ACIS-S and merged
ACIS-I observations is shown in Fig.~\ref{fig:xim_both_obs}. 
Point-like sources identified within each observation are enclosed
by circular or rectangular apertures. The only significant differences
between the ACIS-S and ACIS-I 
images are changes in the brightness of some of the point-like
X-ray sources due to intrinsic variability, and drops in the apparent
diffuse soft X-ray surface brightness in the bottom and top left of the
ACIS-I image due to the gaps between the CCD chips.

Images with the point-sources removed
are shown in panels e to f of  Fig.~\ref{fig:xim_both_obs}, and
were created by replacing events within the
point source regions by a Poissonian
random deviate consistent with the events detected within a 
local background annulus of width $1\farcs5$
around each source-region 
(CIAO task {\sc dmfilth} run in POISSON mode).
Care was taken to ensure that these background regions did not themselves
include any other point source emission. This source removal method is
performed on each image. Images in an energy band $< 3$ keV
have had the point sources removed that were detected only in the soft
$E=0.3$--2.0 keV energy band, as sources detected only
within the hard $E=2.0$--8.0 keV energy band are most likely
heavily obscured and hence invisible at lower energies. 
Hard X-ray images are cleared of
point sources detected in either of the soft or hard X-ray bands.
This method produces a source-subtracted
image of the 33ks ACIS-I data that is essentially identical 
that from the PSF-fitting and subtraction method presented in 
\citet{griffiths2000}.

A more detailed break down of the spatial distribution of the
diffuse X-ray emission within the central 2 kpc of M82
as a function of photon energy 
is shown in Fig.~\ref{fig:xim_sub_bands}, based on the ACIS-S
data alone. For comparison we also plot ground-based optical
R-band and continuum-subtracted \halpha+[\ion{N}{2}] images of
the same region. 

The effect of  absorption by material between us and
in the X-ray emitting plasma near the plane of the galaxy 
is visible at X-ray energies $E< 1.1$ keV as a band of
low X-ray surface-brightness bisecting the
wind from north east to south west, obscuring the nuclear
starburst region. This X-ray 
absorption band has a larger minor-axis extent than
the narrower dust lanes seen in the optical images
shown in Fig.~\ref{fig:xim_sub_bands}. 

At X-ray energies $E\la 0.6$ keV
(at energies around the E=0.57 \ion{O}{7} and 0.65 keV \ion{O}{8} emission
lines)
the central region of the galaxy is heavily obscured, and
the brightness of the diffuse X-ray emission peaks at $z\ga 20\arcsec$ 
($\sim400$ pc) from the mid-plane.
In contrast the mid-plane, nucleus and the inner wind
are visible in emission from
Helium and Hydrogen-like ions of Magnesium ($E\sim 1.3$ -- 1.5 keV,
where the similarity in morphology to the \halpha+[\ion{N}{2}] emission
is striking) and Silicon ($E\sim 1.8$ to 2.0 keV), and the
brightest emission occurs within $z\sim 50$ pc of the mid-plane. 
\citet{read02} and \citet{origlia04}
studied the nuclear-region element abundance 
(within $\sim 15\arcsec$ radius of the center
of the galaxy) using {\it XMM-Newton}
RGS grating data. The \citeauthor{origlia04} study derived a hot-phase 
Oxygen abundance that is unusually low compared
to other $\alpha$-elements.
Neither study could take account of the strong spatial
variations in foreground absorption within this region (because of
the low spatial resolution of {\it XMM-Newton}), which our
{\it Chandra} data shows to be of particular significance in M82 
at energies near the oxygen lines. Fitting simple
models to spectra with intrinsically complex absorption is known
to lead to systematic errors in deriving X-ray abundances 
from CCD-resolution X-ray spectra \citep{weaver00}. We therefor suspect
that the abundances derived from {\it XMM-Newton} studies of M82 
in the \citet{read02} and \citet{origlia04}, in particular the 
apparently low Oxygen abundance, are incorrect.

Diffuse hard X-ray emission is clearly visible within the central
starburst region, even in images without point source subtraction. 
Unlike the diffuse soft X-ray
emission from the superwind, the spatial extent of the brightest
region of diffuse hard X-ray emission is larger along major 
axis (\ie along the plane of the
galaxy) than along the minor axis. Measured by eye,  
hard diffuse X-ray emission
has a major-axis radial extent of $\sim20 \arcsec$ ($\sim 350$ pc) 
approximately centered on the dynamical center of the galaxy, 
and extends $\sim 8 \arcsec$ ($\sim 140$ pc)
and $\sim 12 \arcsec$ ($\sim 210$ pc) above and below the plane of the galaxy 
(to the south east and north west of the plane, respectively).
Note that diffuse hard X-ray emission can be found
outside this central region, although at lower surface brightness. This
is discussed further in \S~\ref{sec:result:images:diff_extent}.

The spatial distribution of the diffuse hard X-ray
emission is relatively smooth. This is not a limitation of spatial
resolution, photon statistics or the method of point source removal, 
as significant structure is visible in the soft X-ray images.
This suggests that the source of the hard X-ray emitting material
is also relatively uniformly distributed within the starburst region.

There are three bright point-like X-ray sources just to the west of the
center of the galaxy in the ACIS-S and ACIS-I observations (one
of which is the variable ultra-luminous X-ray source described in
\citealt{kaaret01,matsumoto01}). Point-source
removal is difficult in this region, as bright sources have large PSF wings.
Combined with the close proximity of these sources  to each other we were
forced to use a $10\farcs8 \times 4\farcs4$ rectangular region
for point source removal. The flux at each pixel within 
this region in the source-subtracted images
is a Poisson random deviate drawn from the distribution
of the fluxes within $1\farcs5$ of the edge of this rectangle, which
includes the region of brightest diffuse hard X-ray emission
at the mid-plane of the disk. This
method is likely to slightly over-estimate the diffuse X-ray surface 
brightness at the off-plane location of the ULX, as this entire region is
filled in by {\sc dmfilth} approximately uniformly. 
Thus the apparent association between the
ULX and a bright region of diffuse hard X-ray emission extending
off the bright mid-plane ridge (\eg seen in
Fig.~\ref{fig:xim_sub_bands}h)
may not be real.

\begin{deluxetable*}{lrrrrrr}
\tablecolumns{6}
\tablewidth{0pc}
\tablecaption{Count rates and diffuse emission fractions.
        \label{tab:diff_frac}}
\tablehead{
\colhead{Energy}
     & \colhead{$R_{\rm tot}$} & \colhead{$R_{\rm diff}$}
     & \colhead{$R_{\rm ps}$} & \colhead{$R_{\rm bg}$}
     & \colhead{$f_{\rm diff}$} & \colhead{$f_{\rm diff}^{T}$}\\
\colhead{(keV)}
     & \colhead{(cts/s)} & \colhead{(cts/s)}
     & \colhead{(cts/s)} & \colhead{(cts/s)}
     & \colhead{\nodata} & \colhead{\nodata} \\
\colhead{(1)}
     & \colhead{(2)} & \colhead{(3)}
     & \colhead{(4)} & \colhead{(5)}
     & \colhead{(6)} & \colhead{(7)}
}
\startdata
\cutinhead{ACIS-S observation (S3 chip)}
  0.3 -- 0.6
    &  $  0.0208\pm{ 0.0012} $
    &  $  0.0191\pm{ 0.0011} $
    &  $  0.0017\pm{ 0.0004} $
    &  $ 0.00089\pm{0.00005} $
    &  $  0.917\pm{ 0.203} $  & $0.91\pm{0.20}$
    \\
  0.6 -- 1.1
    &  $  0.5603\pm{ 0.0057} $
    &  $  0.5263\pm{ 0.0055} $
    &  $  0.0340\pm{ 0.0014} $
    &  $ 0.00051\pm{0.00004} $
    &  $  0.939\pm{ 0.041} $  & $0.93\pm{0.05}$
    \\
  1.1 -- 1.6
    &  $  0.4783\pm{ 0.0052} $
    &  $  0.4347\pm{ 0.0050} $
    &  $  0.0437\pm{ 0.0016} $
    &  $ 0.00030\pm{0.00003} $
    &  $  0.909\pm{ 0.035} $  & $0.90\pm{0.04}$
    \\
  1.6 -- 2.2
    &  $  0.3298\pm{ 0.0044} $
    &  $  0.2727\pm{ 0.0040} $
    &  $  0.0571\pm{ 0.0018} $
    &  $ 0.00057\pm{0.00004} $
    &  $  0.827\pm{ 0.029} $  & $0.81\pm{0.04}$
    \\
  2.2 -- 2.8
    &  $  0.1256\pm{ 0.0027} $
    &  $  0.0901\pm{ 0.0023} $
    &  $  0.0356\pm{ 0.0015} $
    &  $ 0.00025\pm{0.00003} $
    &  $  0.717\pm{ 0.033} $  & $0.69\pm{0.04}$
    \\
  0.3 -- 2.8
    &  $  1.5149\pm{ 0.0093} $
    &  $  1.3429\pm{ 0.0087} $
    &  $  0.1721\pm{ 0.0032} $
    &  $ 0.00252\pm{0.00008} $
    &  $  0.886\pm{ 0.017} $  & $0.87\pm{0.03}$
    \\
  5.0 -- 6.0
    &  $  0.0485\pm{ 0.0017} $
    &  $  0.0138\pm{ 0.0009} $
    &  $  0.0347\pm{ 0.0014} $
    &  $ 0.00037\pm{0.00003} $
    &  $  0.284\pm{ 0.015} $  & $0.20\pm{0.01}$
    \\
  6.0 -- 7.0
    &  $  0.0289\pm{ 0.0013} $
    &  $  0.0079\pm{ 0.0007} $
    &  $  0.0210\pm{ 0.0011} $
    &  $ 0.00041\pm{0.00003} $
    &  $  0.274\pm{ 0.020} $  & $0.19\pm{0.02}$
    \\
  3.0 -- 9.9
    &  $  0.3166\pm{ 0.0043} $
    &  $  0.1071\pm{ 0.0026} $
    &  $  0.2095\pm{ 0.0035} $
    &  $ 0.00548\pm{0.00011} $
    &  $  0.338\pm{ 0.007} $  & $0.26\pm{0.01}$
    \\
\cutinhead{Merged ACIS-I data (I3 chip)}
 0.3 -- 0.6
    &  $  0.0085\pm{ 0.0005} $
    &  $  0.0082\pm{ 0.0005} $
    &  $  0.0004\pm{ 0.0001} $
    &  $ 0.00035\pm{0.00011} $
    &  $  0.956\pm{ 0.288} $  & $0.95\pm{0.23}$
    \\
  0.6 -- 1.1
    &  $  0.3252\pm{ 0.0026} $
    &  $  0.3172\pm{ 0.0026} $
    &  $  0.0080\pm{ 0.0004} $
    &  $ 0.00038\pm{0.00011} $
    &  $  0.975\pm{ 0.053} $  & $0.97\pm{0.06}$
    \\
  1.1 -- 1.6
    &  $  0.3773\pm{ 0.0028} $
    &  $  0.3447\pm{ 0.0027} $
    &  $  0.0326\pm{ 0.0008} $
    &  $ 0.00022\pm{0.00009} $
    &  $  0.914\pm{ 0.025} $  & $0.91\pm{0.04}$
    \\
  1.6 -- 2.2
    &  $  0.2276\pm{ 0.0022} $
    &  $  0.1877\pm{ 0.0020} $
    &  $  0.0398\pm{ 0.0009} $
    &  $ 0.00025\pm{0.00010} $
    &  $  0.825\pm{ 0.021} $  & $0.81\pm{0.03}$
    \\
  2.2 -- 2.8
    &  $  0.0949\pm{ 0.0014} $
    &  $  0.0691\pm{ 0.0012} $
    &  $  0.0258\pm{ 0.0008} $
    &  $ 0.00020\pm{0.00009} $
    &  $  0.728\pm{ 0.024} $  & $0.71\pm{0.03}$
    \\
  0.3 -- 2.8
    &  $  1.0336\pm{ 0.0047} $
    &  $  0.9269\pm{ 0.0044} $
    &  $  0.1067\pm{ 0.0015} $
    &  $ 0.00141\pm{0.00019} $
    &  $  0.897\pm{ 0.013} $  & $0.89\pm{0.03}$
    \\
  5.0 -- 6.0
    &  $  0.0405\pm{ 0.0009} $
    &  $  0.0101\pm{ 0.0005} $
    &  $  0.0305\pm{ 0.0008} $
    &  $ 0.00024\pm{0.00009} $
    &  $  0.249\pm{ 0.009} $  & $0.20\pm{0.01}$
    \\
  6.0 -- 7.0
    &  $  0.0233\pm{ 0.0007} $
    &  $  0.0063\pm{ 0.0004} $
    &  $  0.0170\pm{ 0.0006} $
    &  $ 0.00024\pm{0.00009} $
    &  $  0.269\pm{ 0.013} $  & $0.22\pm{0.01}$
    \\
  3.0 -- 9.9
    &  $  0.2473\pm{ 0.0023} $
    &  $  0.0830\pm{ 0.0014} $
    &  $  0.1644\pm{ 0.0019} $
    &  $ 0.00212\pm{0.00023} $
    &  $  0.335\pm{ 0.005} $  & $0.29\pm{0.01}$
    \\
\enddata
\tablecomments{Column 1: Energy band associated with the following
    count rates.
    Columns 2, 3 and 4: Background-subtracted count rate within
    a $0\farcm476$ radius of 09:55:51.9, +69:40:47.1 (J2000.0). 
    The total background-subtracted count rate (including
    both diffuse emission and point sources) is $R_{\rm tot}$, while
    $R_{\rm diff}$ and $R_{\rm ps}$ are the count rates associated with
    diffuse and point-source emission respectively.
    Column 5: The background count rate within the same region 
    (see \S~\ref{sec:data_analysis:background}).
    Column 6: The apparent 
    diffuse fraction $f_{\rm diff}=R_{\rm diff}/R_{\rm tot}$.
    Column 7. The diffuse fraction after correcting for unresolved point
    sources, assuming $x_{r}=0.90$ for the ACIS-S observation and
    $x_{r}=0.94$ for the merged ACIS-I observation. See 
    \S~\ref{sec:results:images:fdiff} for further details.
    The errors quoted in columns 2 -- 6 are $1\sigma$ statistical 
    uncertainties, while the error quoted in column 7 is the combination
    of the statistical uncertainty in $f_{\rm diff}$ and an assumed 3\%
    uncertainty in $x_{r}$.
    }
\end{deluxetable*}

\subsection{Diffuse emission and unresolved point sources}
\label{sec:results:images:fdiff}

Count rates for the diffuse emission and the detected point sources
within 500 pc of the center of M82 are given in Table~\ref{tab:diff_frac}
for both the ACIS-S and merged ACIS-I observations (the results
for the 33 ks ACIS-I observation are consistent with the merged
observations). The diffuse count rates are based on the point-source
removal technique described above. The point source count rate
within any energy band is the difference between the 
total count rate and the estimated diffuse count rate.

The fraction of the X-ray counts within each energy band from apparently 
diffuse
emission, the diffuse fraction $f_{\rm diff}$, is also presented in
Table~\ref{tab:diff_frac}. Some fraction of the apparently diffuse
emission comes from unresolved point sources, as only the members
of the M82 X-ray source population with individual 
luminosities $L_{\rm X} \ga 4\times10^{36}$
erg s$^{-1}$ are detected as point-like in these observations 
\citep{strickland04a}. Star forming galaxies typically have
X-ray point source populations with cumulative luminosity functions of
the form $N(>L) \propto L^{-\gamma}$, where $\gamma < 1$ 
\citep{colbert04}, \ie the total luminosity is dominated by the
brightest sources. Thus the ratio of the total luminosity in 
detected point sources
to the total luminosity from all point sources is
$x_{r} \approx 1 - (L_{\rm min}/L_{\rm max})^{1-\gamma}$.
Here $L_{\rm min}$ and $L_{\rm max}$ are the luminosities 
of the faintest and brightest point sources detected.
Using this to correct the diffuse fraction for the expected
unresolved point source contribution we obtain 
an estimate of the true diffuse emission fraction
$f_{\rm diff}^{\rm T} = 1 - (1-f_{\rm diff})/x_{r}$.
We used a variety of methods to estimate $x_{r}$ in the ACIS-S
and ACIS-I observations of M82, where the faintest point sources
detected in any energy band typically have $\ga 10$ counts, and
the brightest source is the ULX with several thousand counts. 
We found that for M82 $x_{r}$
depends only weakly on the energy band considered, 
and the largest effect is the assumed luminosity function slope.
We used two values for the slope $\gamma=0.50$  \citep{kilgard03} and 
$\gamma = 0.57$ \citep{colbert04},
deriving estimates of $x_{r}$ in the range 0.87 -- 0.93 for the
ACIS-S observation, and  $x_{r}$ in the range 0.91 -- 0.96 for the
deeper merged ACIS-I observation.

Correcting $f_{\rm diff}$ for unresolved point sources has little effect
at energies below $E\sim 3$ keV, but is a non-negligible correction
at higher energies. Nevertheless, significant residual diffuse hard
X-ray emission exists even after correcting for unresolved point sources,
with a fraction $\ga$ 70 \% ($f_{\rm diff}^{T}/f_{\rm diff}$) 
of the apparent diffuse hard X-ray emission
being unattributable to unresolved point sources.

In general the estimated diffuse fraction is
consistent between the ACIS-S and ACIS-I observations, despite the
differences in spectral sensitivity between the back-illuminated and
front illuminated chips, and the different epochs of observation.
Note that the diffuse fraction is likely to vary with time, in particular in
the harder X-ray bands ($E> 3$ keV) as significant
variability is associated with the nuclear X-ray point sources 
\citep[\eg][]{collura94,ptak99,matsumoto99,matsumoto01,kaaret01}.
Several of the point-like X-ray source did show minor 
variability in flux between
the ACIS-I and ACIS-S observations (as can be seen in 
Fig.~\ref{fig:xim_both_obs}), but this was not of a magnitude sufficient
to significantly alter the diffuse fraction. 

\begin{figure*}[!ht]
\epsscale{1.1}
\plotone{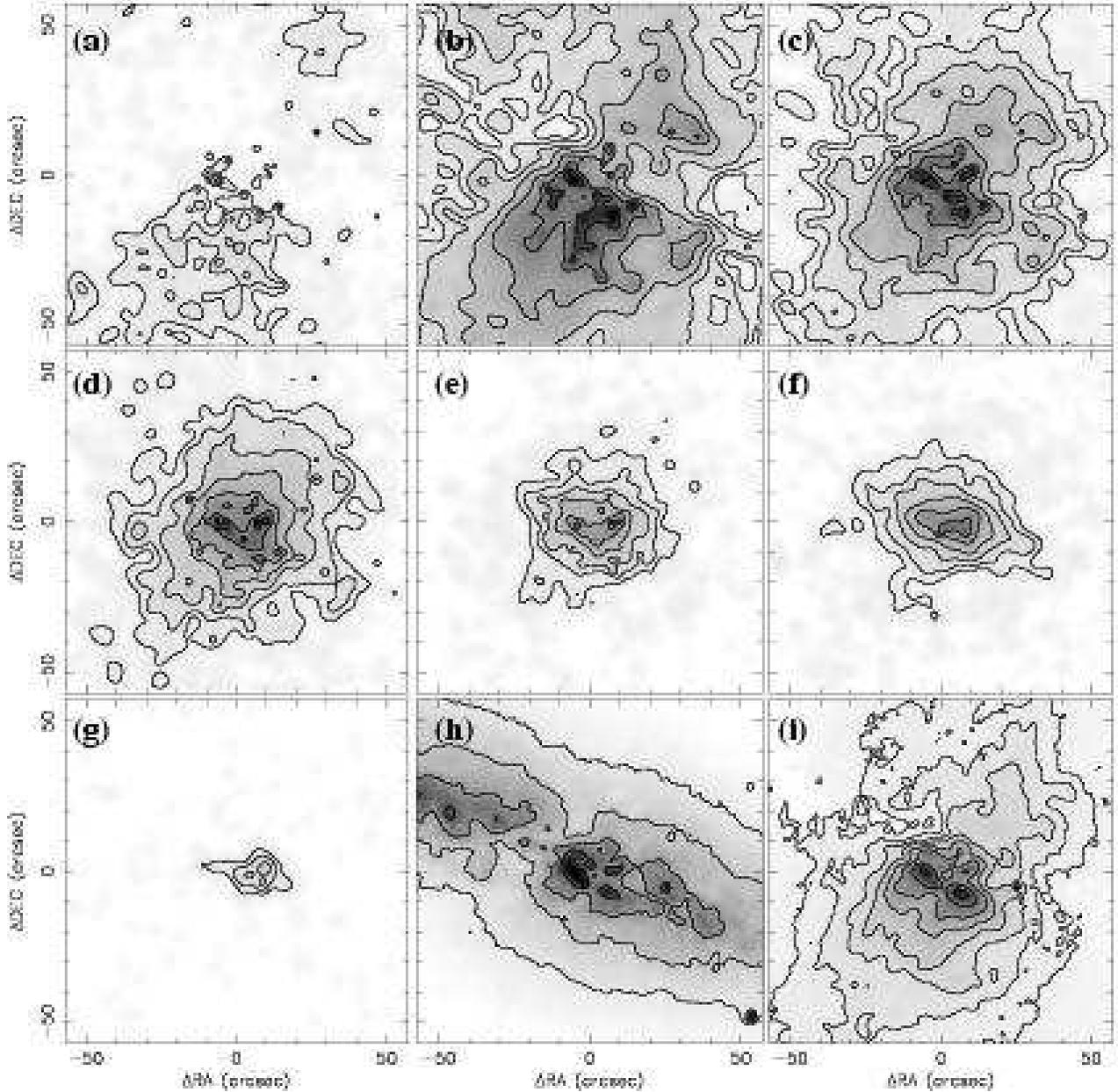}
  \caption{The distribution
  of diffuse X-ray emission in the central $2\times2$ kpc of M82
  as a function of energy is shown in panels a to g, 
  compared to optical R-band continuum and continuum-subtracted
  \halpha+[N II] emission in panels h and i. The energy bands
  used are E=0.3--0.6 keV (a), 0.6--1.1 keV (b), 1.1--1.6 keV (c),
  1.6--2.2 keV (d), 2.2--2.8 keV (e), 3.0--7.0 keV (f) and 6.0-7.0 keV
  (g). 
  All images are shown
  on a square root intensity scale. Contour levels increase in factors of
  2 in surface brightness. Images are aligned so that North is up and
  East is to the left.
  The X-ray images are
  from the ACIS-S observation, have had background and point source
  emission removed, and have been adaptively smoothed so that features are
  locally significant at S/N=3. The intensity scale for the
  X-ray images runs from 0 to 5 counts (per $0\farcs492$ pixel), and
  the contours begin at a level of 0.04 counts per pixel. 
  The effect of strong absorption in the plane of the galaxy 
  is visible at X-ray energies $< 1.1$ keV as a band bisecting the
  wind from north east to south west, obscuring the nuclear
  starburst region.
  }
  \label{fig:xim_sub_bands}
\end{figure*}

The diffuse fraction is high 
($f^{T}_{\rm diff} \ge 70$\%) at energies $E\la 3$ keV, but
is relatively constant with energy at higher energies ($E\ga 3$ keV) at
a value of $f^{T}_{\rm diff} \sim 20$ -- 30\%. 
This transition in diffuse fraction occurs at the
same energy associated with the transition from X-ray emission
extended preferentially along the minor-axis (the soft X-ray emission
in the wind) to predominantly major-axis extended hard diffuse
X-ray emission shown in Fig.~\ref{fig:xim_sub_bands}.

Although the ULX is the single brightest X-ray point source 
over the 0.3--8.0 keV
energy band, it accounts for somewhat less than half of the total resolved
point source related flux from the central kiloparsec of 
M82 in either Chandra ACIS observation.
In the merged ACIS-I observations of 1999 September 20, the ULX
contributed $\sim 51$\% of the 0.3--2.8 keV energy band count rate, 
$\sim 41$\% of the 3.0--9.9 keV band count rate and $\sim45$\%
of the 0.3-9.9 keV band count rate within a radius of $29\arcsec$
of the center of the galaxy. In the ACIS-S observation of 2002 June 18
the corresponding numbers are 19, 39 and 30\% respectively.
We estimate that
the true count rate of the ULX is probably 50\% higher than 
the observed count rate in the soft energy band, and 20\% higher in the hard
energy band, due to the effects of pile-up in the ACIS CCD detectors
(based on the PIMMS count rate prediction tool), but this does
not alter the general conclusion that in total the other resolved point sources
contribute a comparable flux to that of the ULX at the specified epochs.

\subsection{The spatial extent of the diffuse hard X-ray emission}
\label{sec:result:images:diff_extent}

\begin{figure*}[!ht]
\epsscale{1.1}
\plotone{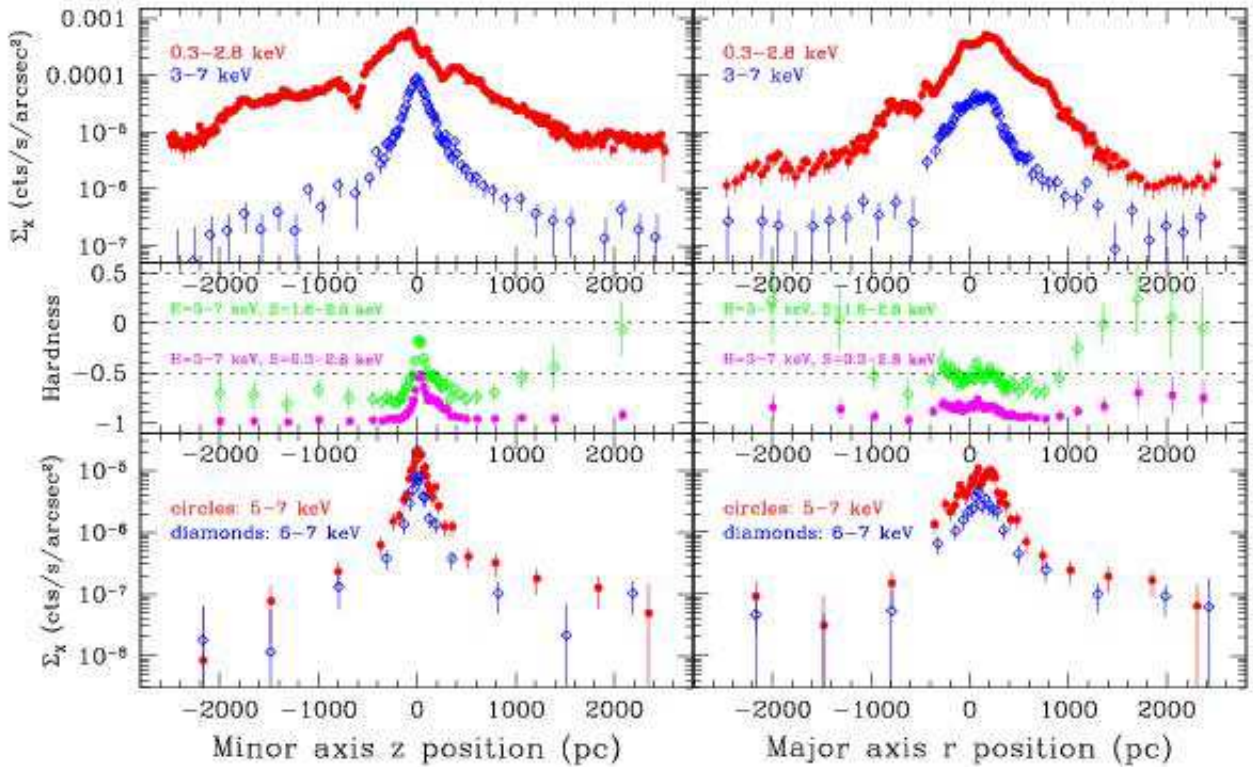}
  \caption{Minor and major axis diffuse X-ray emission
        surface brightness and spectral
	hardness profiles, from the merged ACIS-I observation. The positive
        $z$ axis is along $PA=343\degr$, 
        and the positive $r$ axis along $PA=253\degr$.
        The background  (\S~\ref{sec:data_analysis:background})
        has been removed, and point sources interpolated over.
        Error bars are $1\sigma$ uncertainties. The spectral hardness ratio
        is defined as $Q=(H-S)/(H+S)$, 
        where $H$ and $S$ are the counts in the higher and
        lower energy bands respectively. 
	}
  \label{fig:profile_acisi}
\end{figure*}

We created surface brightness profiles along the major and minor axes
of the galaxy in several different energy bands in order
to better quantify the spatial extent of the diffuse hard X-ray emission 
in M82, particularly in comparison to the better studied 
diffuse soft X-ray emission. We also calculated the spectral
hardness $Q=(H-S)/(H+S)$, where $H$ and $S$ are the counts in chosen
hard and soft energy bands respectively, along these axes.
Only data within ${48}\arcsec$ (${835}$ pc) of the respective axis
was used when creating each profile in order to achieve a
reasonable balance between maximizing the number of counts without
averaging over too large an area.
We re-binned each profile so as to achieve $\ge 20$ counts per bin
per energy band
for plotting and fitting the surface brightness profiles, and 
$\ge 40$ counts per bin per energy band when calculating the spectral
hardness.
The resulting profiles for the merged ACIS-I observation is are shown in 
Fig.~\ref{fig:profile_acisi}. The ACIS-S surface brightness profiles 
are in general very similar to the ACIS-I profiles shown in this
figure. The most notable differences
are purely instrumental due to the effects of the 
gaps between the ACIS-I CCD chips 
(at $z\sim -600$ pc and $r \sim -400$ pc). At photon energies $E \ga 1$ keV
all profiles show a central peak in emission, the surface brightness
of which drops off rapidly with distance, surrounded by 
more-extended lower surface brightness X-ray emission. 

Along both major and minor axes the hardness ratios show that the emission
within $\la 400$ pc of the center of the galaxy is spectrally distinct from the
kiloparsec scale emission. The effect of absorption by the disk
is seen most clearly in the minor axis hardness ratios.
Along the minor axis this hardness ratio drops
from the nuclear starburst region to the larger scale wind. Moving to
the SE along the wind the hardness is relatively constant in
both the ACIS-S and ACIS-I data, although in contrast the hardness ratio
increases moving NW from the nucleus. 

Simple model fitting demonstrates that to first order
the scale heights of the soft and hard diffuse X-ray surface 
brightness outside the nuclear region are 
very similar, $H \sim 500$ -- 1000 pc.
A particularly interesting result is
the apparent detection of
diffuse hard X-ray emission out to larger $z$ and $r$ distances 
than recognized in the \citet{griffiths2000} analysis of the initial
ACIS-I observations 
(out to $|z| \sim |r| \sim 2$ kpc). However this  faint extended  
hard emission X-ray might
be due to the extended wings of the Chandra 
point spread function, rather than being genuinely 
diffuse emission. We will investigate this issue further in the
future.

\subsection{Comparison to features at other wavelengths}

It is instructive to compare the spatial distribution of the 
soft and hard X-ray within the central region of M82 to
starburst-related activity seen at other wavelengths. In 
Fig.~\ref{fig:xray_radio_comp} we plot the location
of the near-IR super-star clusters
\citep{mcgrady03}, the radio SNRs \citep{unger84,kronberg85,muxlow94,wills98}, 
and the expanding shells seen described in \hi~\citep{wills02}
and CO-emission\footnote{\citet{wills02} show that
their shell 3 is most-probably the
northern side of the CO-superbubble discussed in
\citet{weiss99} and \citet{matsushita00}, the 
exact energetics and mass of which
are not agreed upon.} \citep{neininger98,weiss99,matsushita00}.

\begin{figure*}[!ht]
\epsscale{1.1}
\plotone{f4.jps}
  \caption{
  The X-ray emission from the central $1.5\times1.5$ kpc region of M82 
  is shown in comparison to the near-IR super-star clusters of
  \citet[small blue crosses]{mcgrady03}, the radio SNRs of 
  \citet[red X's]{wills98}, the \hi~shells described in 
  \citet[shown as blue circles
  of diameter equivalent to the observed shell]{wills02}
  and the CO-superbubble \citep[cyan polygon]{weiss99}. 
  Panels a and b show ACIS-S images in the $E=0.3$--2.0 keV energy band,
  including the point source emission in (a) and source-subtracted in (b).
  Adaptive smoothing has been used in (a), whereas in (b) the smoothing
  used a Gaussian mask of FWHM$=2\arcsec$. Panel (c) is equivalent to (a),
  but shows the merged ACIS-I image.
  Panels (d) through (f) are equivalent to panels (a) through (c), except
  the images are in the 2.0 -- 8.0 keV energy band.
  }
  \label{fig:xray_radio_comp}
\end{figure*}

\citet{matsumoto01} discuss the relationship between the X-ray
point sources seen in the {\it Chandra} High Resolution Camera 
observations of M82 and the 23 radio sources seen at 5 GHz presented
in \citet{muxlow94}. 
The coordinates of the X-ray point sources
seen in both our re-processed ACIS observations and the published
HRC observations match within the uncertainties. 
In Fig.~\ref{fig:xray_radio_comp} we plot the locations of the 33 compact
radio sources observed at 1.4 GHz by \citet{wills98}.
The clearest association of an X-ray source with a 
radio source at either 1.4 or 5 GHz is with the enigmatic radio source 41.9+58,
one of the few radio sources in M82 that is fading with time 
\citep{kronberg00}. The radio source 44.0+59.6, another of the brightest
radio sources, falls at the location of a weaker hard X-ray point source.
In general the radio sources, most of which are young SNRs evolving
in confining dense gaseous environments \citep{kronberg00,chev01}, 
are not associated with
X-ray point sources.

\begin{figure*}
\epsscale{1.1}
\plotone{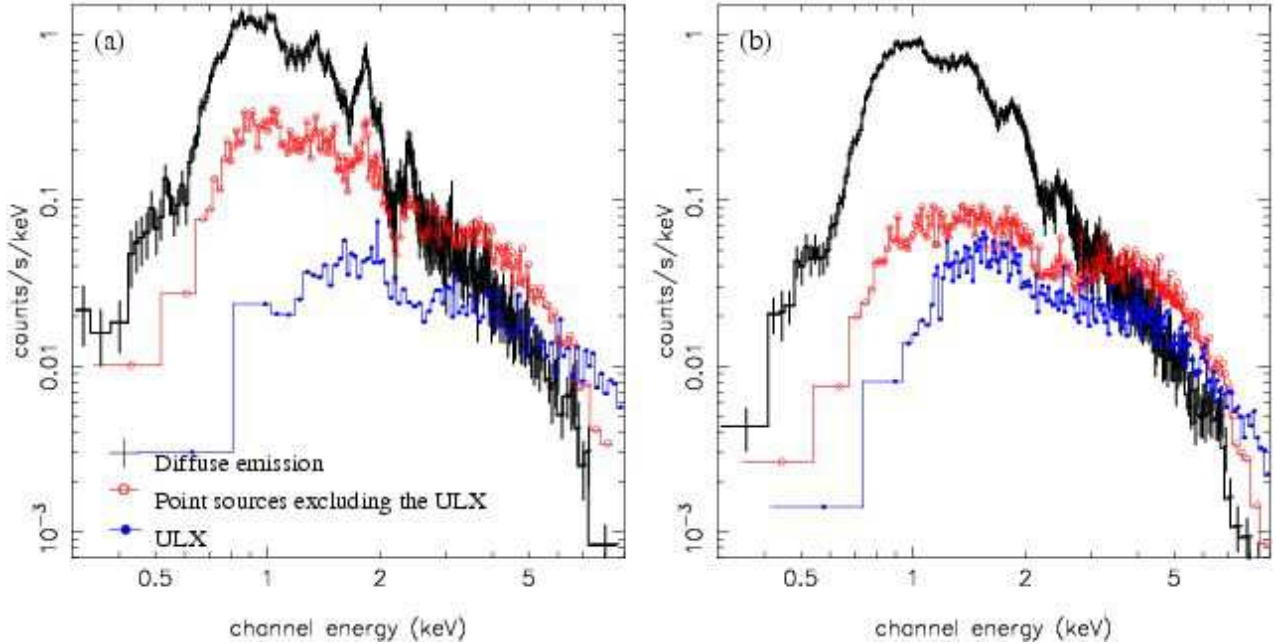}
  \caption{Background-subtracted ACIS spectra of the nuclear
	diffuse emission, the summed nuclear point sources (excluding
	the ULX), and the ULX itself as seen in the ACIS-S observation
	(a) and the merged ACIS-I observations (b). To improve the
	clarity of the figure the error bars are not shown on the 
	point source and ULX spectra. The primary
	differences between
	the ACIS-S and ACIS-I spectra are purely instrumental, and are
	due to the poorer low energy
	sensitivity and spectral resolution of the cosmic-ray-damaged
	ACIS-I detectors.
  }
  \label{fig:raw_spectra}
\end{figure*}

Although the SNRs associated with radio sources may not directly 
contribute to the mechanical powering of the superwind because of
their confinement, they must delineate the region associated with 
current SN activity.
It is noteworthy that the major-axis extent of both the soft and hard
diffuse X-ray emission is so similar to that of the radio sources ---
this is the unambiguous sign of a massive-star-related SN-driven wind.
The distribution of the diffuse hard X-ray emission differs from that
of the radio sources only in that it has a larger minor-axis extent,
to be expected if the X-ray-emitting plasma is being advected out along
the minor axis. 

The near-IR star clusters defined in \citet{mcgrady03} are more widely
distributed along the major axis of the galaxy than the radio sources
or the soft and hard diffuse X-ray emission. However not all
of these clusters are young enough to support ongoing core-collapse
SN activity. It is not surprising that
there is very little
diffuse X-ray emission in the vicinity of the $60\pm{20}$ Myr-old super
star cluster F \citep{smith01}. One of the point-like X-ray sources
may be associated with the nearby cluster L.

Three of the four expanding \hi~shells identified by \citet{wills02} lie
within the region covered by diffuse soft and hard X-ray emission. As
mentioned previously, one of these \hi~shells is probably associated with
the CO-bubble discussed in \citet{weiss99} and \citet{matsushita00}.
This entire region along the major axis 
is the location of the brightest soft and hard diffuse
X-ray emission, there being no discernible difference in X-ray surface
brightness between the regions interior and exterior to the \hi~and/or
CO shells.

\section{Spectral analysis of the Chandra and XMM-Newton data}
\label{sec:results:spectra}

We will begin with an analysis of the {\it Chandra} ACIS-S and ACIS-I 
spectra of each of the diffuse, point source and ULX spectra, followed by
assessment of the significance of the $E \sim 6.4$ keV and $E \sim 6.7$ 
keV iron 
line features in those spectra. We will then present a more
elaborate model that attempts to account for the presence of unresolved 
point source emission in the diffuse spectrum, and assess what
limits can be placed on $E=6.97$ keV emission from hydrogen-like 
ionized iron in the {\it Chandra} spectra. This section is
concluded with analysis of the integrated nuclear spectra
as seen in the two separated {\it XMM-Newton} observations 
(by integrated we mean that the spectra
include diffuse, point source and ULX emission). We will
show that a full picture of the origin of the 
iron line emission from M82 is only obtained if the
data from both {\it Chandra} and {\it XMM-Newton} at
all epochs is assessed.

The Chandra ACIS spectra of the central kiloparsec of M82 are quite complex,
particularly in the soft X-ray band, where spatially-varying absorption
and possible variations in mean plasma temperature occur 
(see Fig.~\ref{fig:raw_spectra}). With
the $\Delta E \sim 0.1$ keV resolution of the ACIS-I and ACIS-S CCD detectors
the strong line emission from O, Fe, Ne, Mg, Si, and  S are heavily
blended, making a unique and physically accurate spectral fit currently
difficult. Our focus in this paper is instead 
on the harder X-ray emission with energies $E\ga 2$ keV, where
there are fewer spectral features and absorption can effectively be
ignored. In particular we are interested in the continuum spectral
shape of the diffuse hard X-ray emission, and whether it is similar
to that of the resolved point source population. 
Given the ambiguous detection of iron line emission in the
\citet{griffiths2000} study, 
another question
we wish to answer is what are the statistical significances and fluxes
of any $E \sim 6.4$, 6.7 or 6.9 keV iron line emission from
the diffuse component? Is the
Fe-K emission associated with the point source population, and/or the
enigmatic luminous and variable ULX, as \citet{strohmayer03} 
conclude based on the ObsID 112290201 {\it XMM-Newton} observation of M82?

\subsection{Chandra ACIS spectra}
\label{sec:results:spectra:chandra}

We extracted spectra from the ACIS-S and ACIS-I datasets within
a radius of $28\farcs6$ (equivalent to a physical radius of 500 pc) 
of the dynamical center of M82 at $\alpha=09^{\rm h} 55^{\rm m} 51\fs9, 
\delta=+69\degr 40\arcmin 47\farcs1$ (J2000.0). 
We extracted and fit spectra to both the merged ACIS-I observation 
(which has the greatest number of counts) and the 33ks ACIS-I observation
(the observation used by \citealt{griffiths2000}).
A spectrum of the diffuse X-ray emission was created by 
excluding all detected point sources using either circular
or rectangular masks, as shown in Fig.~\ref{fig:xim_both_obs}. 
The source extraction radius used depended on
source flux, and ranged from $1\farcs2$ to $3\farcs0$, most typically
being $1\farcs5$ -- $2\arcsec$. These radii were chosen to maximize
the removal of point-source related flux from the diffuse spectrum: a 
radius of $1\farcs5$
encloses $\sim 85$ -- 97\% of the flux from a point source, depending
on the energy of the photons (the enclosed energy fraction is smaller at
higher energies). We also created a summed spectrum of the point source
population (excluding the ULX) using these point source extraction regions,
and a spectrum of the ULX itself.  A small component of the point 
source and ULX spectra is from truly diffuse emission within the extraction
regions. The spectrum of the ULX suffers from ``pile-up,''
in which two or more
X-ray photons arrive in a single pixel within a single 3.2s ACIS
exposure and are falsely detected as a single photon of higher energy.
We do not seek to correct for this process, but merely to see if any Fe-K
line emission can be detected above the local continuum.

We only fit the ACIS spectra in the energy range
$E=3.15$ --3.80 keV and $4.20$ -- 9.0 keV. Excluding the data below 
3.15 keV removes any contribution
from the soft diffuse emission and associated Si and S emission, as well
as the instrumental Ir edges.
We exclude the energy range 3.8-4.2 keV to avoid any He-like and H-like Calcium
emission, which is present at marginal significance in the ACIS-S diffuse
emission spectrum but is not obvious in the ACIS-I spectra. Our aim is 
to simplify the spectral fitting by essentially only considering a
pure continuum plus any iron lines. 

When modeling the continuum in the diffuse and point source
spectra we found that 
standard thermal bremsstrahlung and power law models were adequate.
The spectra of the ULX are heavily piled up, and can not be adequately
fit with power law or bremsstrahlung models (even when we also included the 
``pileup'' model available in XSPEC). We found that a simple broken power
law model, with the transition between the two power law slopes occurring
at $E\sim 6.0\pm0.5$ keV, 
provided the best fit the the apparent continuum in the ULX spectra.
No absorption model was used, as a column $\nH \sim 10^{23}$ is required
to produce an optical depth of $\tau \sim 1$ at $E\sim 3$ keV.

At the spectral resolution of ACIS-S and ACIS-I the Fe-K$\alpha$ fluorescence
lines from nearly neutral iron and  the Fe He$\alpha$ line complex from highly
ionized iron can be modeled as narrow Gaussian ($\sigma=1$ eV) lines
at $E=6.40$ and $E=$6.69 keV respectively.
Although it is possible to fit the ACIS spectra to determine the 
the line energy and/or width this does
not place any useful constraint on these parameters, nor does this
result in statistically superior fits compared to models using the expected
line width and energy (by statistically superior we mean a change in
chi-squared per change in degrees of freedom 
$\Delta \chi^{2}/\Delta \nu \ge 1$). For example,
when fitting for the line energy of the Fe He$\alpha$ line the best fits 
were $E=6.66^{+0.10}_{-0.11}$
keV (ACIS-S) and $E=6.69\pm{0.17}$ keV (merged ACIS-I spectrum).
Thus for the majority of the spectral fits described below, the
line energies and widths were fixed at the expected values, 
and we fit only for each line intensity.


\begin{deluxetable*}{lrllrrrrrrrr}
\tablecolumns{12}
\tablewidth{0pc}
\tablecaption{Model fits to the diffuse, point source and ULX spectra
        \label{tab:fits}}
\tablehead{
\colhead{Model} & \colhead{Dataset} 
    & \multicolumn{2}{l}{Continuum model}
    & \multicolumn{2}{l}{6.4 keV Fe K$\alpha$ line}
    & \multicolumn{2}{l}{6.7 keV Fe He$\alpha$ line} 
    & \multicolumn{2}{l}{6.9 keV Fe Ly$\alpha$ line} 
    & \colhead{$\chi^{2}$} & \colhead{$\nu$} \\
\colhead{} & \colhead{} 
    & \colhead{$kT$ or $\Gamma$} & \colhead{norm}  
    & \colhead{norm}  & \colhead{EW}
    & \colhead{norm}  & \colhead{EW}  
    & \colhead{norm}  & \colhead{EW}  
    & \colhead{} & \colhead{} \\  
\colhead{(1)} & \colhead{(2)} 
    & \colhead{(3)}  & \colhead{(4)} 
    & \colhead{(5)} & \colhead{(6)}
    & \colhead{(7)} & \colhead{(8)} 
    & \colhead{(9)} & \colhead{(10)}
    & \colhead{(11)} & \colhead{(12)}
}
\startdata
\cutinhead{{\it Chandra} diffuse emission spectra}
BR, two lines & ACIS-S
  & $6.2^{+4.7}_{-2.0}$ & $9.7^{+2.8}_{-1.9}$ 
  & $4.4^{+4.6}_{-4.4}$ & 112
  & $8.2^{+5.9}_{-5.9}$ & 273
  & \nodata & \nodata
  & 36.1 & 81 \\
\nodata & 33ks ACIS-I
  & $3.5^{+1.0}_{-0.7}$ & $13.0^{+3.5}_{-2.7}$ 
  & $5.2^{+4.9}_{-4.9}$ & 190
  & $6.3^{+5.4}_{-5.4}$ & 270
  & \nodata & \nodata
  & 65.1 & 115 \\
\nodata & merged ACIS-I
  & $3.4^{+0.7}_{-0.5}$ & $13.6^{+3.0}_{-2.4}$ 
  & $5.1^{+4.0}_{-4.0}$ & 175 
  & $7.7^{+4.6}_{-4.6}$ & 335
  & \nodata & \nodata
  & 94.4 & 136 \\
BR, no lines & ACIS-S
  & $8.0^{+8.1}_{-2.8}$ & $8.8^{+2.2}_{-1.3}$ 
  & \nodata & \nodata
  & \nodata & \nodata
  & \nodata & \nodata
  & 43.5 & 83 \\
\nodata & 33ks ACIS-I
  & $4.4^{+1.3}_{-0.9}$ & $10.9^{+2.6}_{-1.9}$ 
  & \nodata & \nodata
  & \nodata & \nodata
  & \nodata & \nodata
  & 77.1 & 117 \\
\nodata & merged ACIS-I
  & $4.3^{+1.0}_{-0.7}$ & $11.3^{+2.2}_{-1.7}$ 
  & \nodata & \nodata
  & \nodata & \nodata
  & \nodata & \nodata
  & 115.2 & 138 \\
PL, two lines & ACIS-S
  & $2.1^{+0.3}_{-0.3}$ & $13.7^{+8.7}_{-5.3}$ 
  & $4.2^{+4.6}_{-4.2}$ & 108
  & $7.9^{+5.8}_{-5.8}$ & 259
  & \nodata & \nodata
  & 34.8 & 81 \\
\nodata & 33ks ACIS-I
  & $2.7^{+0.3}_{-0.3}$ & $26.4^{+12.8}_{-8.5}$ 
  & $4.9^{+4.9}_{-4.9}$ & 181
  & $5.6^{+5.3}_{-5.4}$ & 232
  & \nodata & \nodata
  & 61.0 & 115 \\
\nodata & merged ACIS-I
  & $2.7^{+0.2}_{-0.2}$ & $28.4^{+10.9}_{-7.8}$ 
  & $4.8^{+4.0}_{-4.0}$ & 165
  & $6.9^{+4.6}_{-4.6}$ & 293
  & \nodata & \nodata
  & 87.2 & 136 \\
PL, no lines & ACIS-S
  & $2.0^{+0.3}_{-0.3}$ & $11.1^{+6.6}_{-4.1}$ 
  & \nodata & \nodata
  & \nodata & \nodata
  & \nodata & \nodata
  & 41.9 & 83 \\
\nodata & 33ks ACIS-I
  & $2.5^{+0.3}_{-0.2}$ & $20.2^{+8.9}_{-6.1}$ 
  & \nodata & \nodata
  & \nodata & \nodata
  & \nodata & \nodata
  & 71.4 & 117 \\
\nodata & merged ACIS-I
  & $2.5^{+0.2}_{-0.2}$ & $21.3^{+7.5}_{-5.5}$ 
  & \nodata & \nodata
  & \nodata & \nodata
  & \nodata & \nodata
  & 105.4 & 138 \\

BR+PL, three lines & ACIS-S
  & $3.8^{+2.5}_{-1.2}$ & $10.3^{+5.4}_{-3.1}$ 
  & $4.4^{+4.6}_{-4.4}$ & 115
  & $8.1^{+5.8}_{-5.8}$ & 253
  & $1.2^{+4.8}_{-1.2}$ & 37
  & 35.1 & 80 \\
\nodata & 33ks ACIS-I
  & $2.8^{+0.8}_{-0.5}$ & $14.9^{+5.5}_{-3.7}$ 
  & $5.4^{+5.2}_{-4.9}$ & 202
  & $5.8^{+5.7}_{-5.5}$ & 235
  & $0.4^{+5.2}_{-0.4}$ & 20
  & 63.1 & 114 \\
\nodata & merged ACIS-I
  & $2.7^{+0.6}_{-0.4}$ & $15.7^{+4.6}_{-3.3}$ 
  & $5.1^{+4.3}_{-4.0}$ & 177 
  & $7.4^{+4.5}_{-5.4}$ & 319
  & $0.0^{+4.1}_{-0.0}$ & 0
  & 91.0 & 135 \\

PL+PL, three lines & ACIS-S
  & $2.5^{+0.5}_{-0.5}$ & $19.1^{+17.8}_{-9.3}$ 
  & $4.2^{+4.6}_{-4.2}$ & 108
  & $7.8^{+5.8}_{-5.8}$ & 242
  & $0.8^{+4.8}_{-0.8}$ & 24
  & 34.3 & 80 \\
\nodata & 33ks ACIS-I
  & $3.0^{+0.3}_{-0.3}$ & $35.1^{+0.4}_{-12.2}$ 
  & $4.7^{+4.4}_{-4.7}$ & 172
  & $5.5^{+5.0}_{-5.5}$ & 227
  & $0.0^{+4.1}_{-0.0}$ & 0
  & 59.9 & 114 \\
\nodata & merged ACIS-I
  & $3.0^{+0.3}_{-0.2}$ & $37.0^{+20.4}_{-9.9}$ 
  & $4.5^{+4.3}_{-3.6}$ & 154
  & $6.8^{+4.3}_{-5.0}$ & 284
  & $0.0^{+3.1}_{-0.0}$ & 0
  & 85.1 & 135 \\

\cutinhead{{\it Chandra} summed point source spectra, excluding the ULX}
PL, two lines & ACIS-S
  & $1.0^{+0.2}_{-0.2}$ & $5.4^{+2.1}_{-1.5}$ 
  & $3.2^{+6.6}_{-3.2}$ & 40
  & $1.1^{+1.1}_{-7.1}$ & 13
  & \nodata & \nodata
  & 83.0 & 134 \\
\nodata & 33ks ACIS-I
  & $1.0^{+0.2}_{-0.1}$ & $4.0^{+1.3}_{-0.7}$ 
  & $0.0^{+3.8}_{-0.0}$ & 0
  & $3.0^{+4.6}_{-3.0}$ & 54
  & \nodata & \nodata
  & 109.5 & 166 \\
\nodata & merged ACIS-I
  & $1.0^{+0.1}_{-0.1}$ & $3.9^{+0.9}_{-0.7}$ 
  & $0.0^{+2.9}_{-0.0}$ & 0
  & $4.5^{+5.4}_{-4.5}$ & 78
  & \nodata & \nodata
  & 130.9 & 206 \\
PL, no lines & ACIS-S
  & $1.0^{+0.2}_{-0.2}$ & $5.2^{+1.9}_{-1.4}$ 
  & \nodata & \nodata
  & \nodata & \nodata
  & \nodata & \nodata
  & 83.7 & 136 \\
\nodata & 33ks ACIS-I
  & $1.1^{+0.2}_{-0.2}$ & $4.1^{+1.1}_{-0.9}$ 
  & \nodata & \nodata
  & \nodata & \nodata
  & \nodata & \nodata
  & 109.3 & 168 \\
\nodata & merged ACIS-I
  & $1.0^{+0.1}_{-0.1}$ & $3.7^{+8.0}_{-6.6}$ 
  & \nodata & \nodata
  & \nodata & \nodata
  & \nodata & \nodata
  & 132.5 & 208 \\
\cutinhead{{\it Chandra} ULX spectra}
BPL, two lines & ACIS-S
  & $0.3^{+0.4}_{-0.3}$, $-4.5^{+0.9}_{-1.5}$ & $0.7^{+0.8}_{-0.4}$ 
  & $1.2^{+6.6}_{-1.2}$ & 19
  & $0.4^{+7.6}_{-0.4}$ & 5
  & \nodata & \nodata
  & 48.9 & 89 \\
\nodata & 33ks ACIS-I
  & $0.6^{+0.3}_{-0.3}$, $-4.2^{+1.1}_{-1.2}$  & $1.1^{+0.8}_{-0.4}$ 
  & $1.7^{+5.4}_{-1.7}$ & 37
  & $1.2^{+4.1}_{-1.2}$ & 23
  & \nodata & \nodata
  & 71.5 & 122 \\
\nodata & merged ACIS-I
  & $0.5^{+0.3}_{-0.2}$, $-3.0^{+0.8}_{-1.4}$ & $1.0^{+0.5}_{-0.3}$ 
  & $0.0^{+4.2}_{-0.0}$ & 0
  & $0.0^{+6.4}_{-0.0}$ & 0
  & \nodata & \nodata
  & 103.3 & 175 \\
BPL, no lines & ACIS-S
  & $0.3^{+0.4}_{-0.4}$, $-4.5^{+0.6}_{-0.8}$ & $0.7^{+0.8}_{-0.7}$ 
  & \nodata & \nodata
  & \nodata & \nodata
  & \nodata & \nodata
  & 49.0 & 91 \\
\nodata & 33ks ACIS-I
  & $0.6^{+0.3}_{-0.3}$, $-2.6^{+0.9}_{-0.9}$ & $1.2^{+0.8}_{-0.5}$ 
  & \nodata & \nodata
  & \nodata & \nodata
  & \nodata & \nodata
  & 71.5 & 124 \\
\nodata & merged ACIS-I
  & $0.5^{+0.2}_{-0.2}$, $-3.1^{+1.0}_{-1.0}$ & $1.1^{+0.5}_{-0.3}$ 
  & \nodata & \nodata
  & \nodata & \nodata
  & \nodata & \nodata
  & 103.3 & 177 \\
\cutinhead{{\it XMM-Newton} nuclear region spectra including point sources and the ULX}
BPL, three lines & XMM-short
  & $0.7^{+0.1}_{-0.1}$, $2.0^{+0.1}_{-0.1}$ & $12.6^{+0.6}_{-0.2}$ 
  & $6.6^{+3.7}_{-3.7}$ & 23
  & $15.0^{+3.9}_{-2.0}$ & 58
  & $6.9^{+3.8}_{-3.9}$ & 28
  & 1024.9 &  1555 \\
\nodata & XMM-long
  & $1.6^{+0.1}_{-0.1}$, $2.6^{+0.2}_{-0.2}$ & $24.7^{+1.8}_{-1.1}$ 
  & $1.3^{+1.5}_{-1.3}$ & 9
  & $8.1^{+1.7}_{-1.6}$ & 70
  & $0.2^{+1.5}_{-0.2}$ & 2
  & 1194.6 & 1573 \\
\enddata
\tablecomments{All errors are 90\% confidence for 1 interesting parameter.
  Column 1: Abbreviations for the continuum model used, and
  whether the 6.4 and 6.7 keV lines were fit for. 
  Thermal Bremsstrahlung: BR; Power Law: PL; Broken
  power law: BPL; Bremsstrahlung with additional power law component
  representing unresolved X-ray binary emission: BR+PL; Power law
  with additional power law component
  representing unresolved X-ray binary emission: PL+PL;
  The break energy for the BPL model is not tabulated,
  but ranged from 5.6 to 6.3 keV in the best fit models to both the
  {\it Chandra} and {\it XMM-Newton} spectra. 
  The power law component used in the BR+PL and PL+PL models to represent
  the unresolved X-ray binary emission is described in 
  \S~\ref{sec:results:spectra:newchandra}.
  Column 3:  Temperature kT (keV) for bremsstrahlung models, or
    photon index for power law models. Two photon indexes are
    given for the broken power law models fit to the ULX.
  Column 4: Continuum model normalization factor.
    These values are numerically $10^{4} \times K$, where $K$  is 
    the XSPEC model normalization factors. For the bremsstrahlung models
    $K=3.02 \times 10^{-15}/4\pi D^{2} \times  \int  n_{e} n_{I} dV$
    where $n_{e}$ and $n_{I}$ are the electron and ion number densities 
    respectively (cm$^{-3}$), $D$ is the distance to the source in cm
    and $V$ the volume of the emitting region.
    For the power law and broken power law models
   $K$ is the photon flux per unit energy 
    (units of photons/keV/cm$^{2}$/s), evaluated at $E=1$ keV 
  Columns 5, 7 and 9: Line normalizations, effectively total line flux
    in units of $10^{-6}$ photons s$^{-1}$ cm$^{-2}$. 
  Columns 6, 8 and 10: Line equivalent widths corresponding to the best-fit
   line and continuum normalizations, in units of eV. 
  Column 11: Best-fit $\chi^{2}$ value.
  Column 12: Total number of degrees of freedom.
    }
\end{deluxetable*}


\begin{figure*}
\epsscale{.9}
\centerline{
\plotone{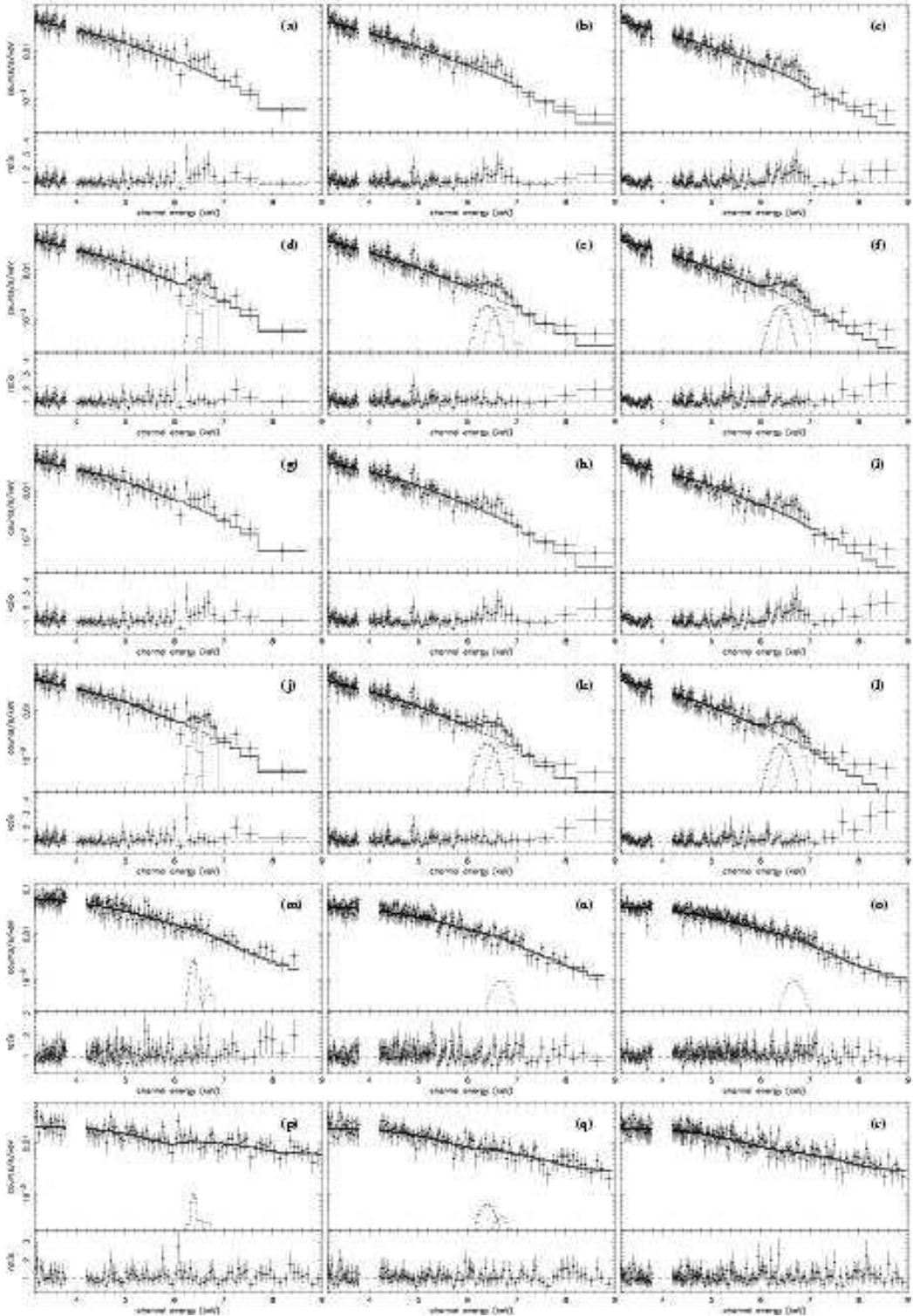}}\epsscale{1}
  \caption{Nuclear diffuse (panels a -- l), point source 
  (excluding the ULX, panels m -- o) and ULX
  spectra (panels p -- r) and best-fit models with residuals. The first column
  are ACIS-S spectra, the second column spectra from the 33ks ACIS-S
  observation and the third column spectra from the merged ACIS-I
  observation. Panels a -- c and g -- i (the first and third rows) 
  are pure power law and
  bremsstrahlung models respectively, in which residuals
  in 6.2 -- 6.8 keV energy range are clearly visible. Panels d -- f and
  j -- l (second and fourth rows) 
 are power law with lines and bremsstrahlung with lines
  model fits to the diffuse spectra. 
  The point source spectra are shown with best-fit models
  of a power law plus lines in panels m -- o, and the ULX spectra
  with a broken power law plus lines model in panels p -- r. The 
  emission lines
  in the point source and ULX models are not required statistically.
  }
  \label{fig:fitted_spectra}
\end{figure*}

\begin{figure*}
\epsscale{1.1}
\centerline{
\plotone{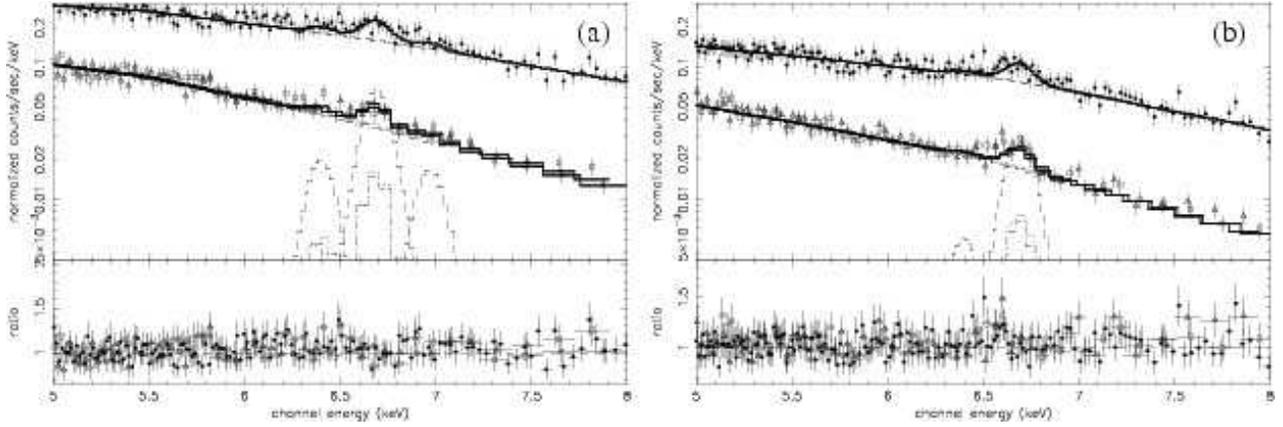}}
  \caption{{\it XMM-Newton} spectra of the nuclear region of M82, along
  with best-fitting models. The shorter of the two observations is
  shown in panel a, and the longer in panel b. Data from the PN detector
  is shown as filled black circles, MOS1 data as open triangles
  and MOS2 data as open squares. The solid, dashed and dotted lines are
  the best fit model components.}
  \label{fig:xmmspectra}
\end{figure*}


The best-fit spectral models are plotted along with the diffuse, point
source and ULX spectra in Fig.~\ref{fig:fitted_spectra}, and the best-fit
model parameters are shown in Table~\ref{tab:fits}. The derived fluxes
for the diffuse emission are shown in Table~\ref{tab:fluxes}.
The pure continuum model fits to the diffuse emission spectra (\ie without any
line components) show strong residuals in the
energy range 6.2 -- 6.8 keV in the ACIS-S, 33ks ACIS-I and merged
(48ks) ACIS-I spectra, and the addition of the two lines 
does a good job of removing these residuals. 
The addition of the Gaussian model components representing Fe K$\alpha$
and Fe He$\alpha$ emission improves the quality of fit in all the diffuse
emission spectra, but not in the point source and ULX spectra. Quantitatively
the  best-fit $\chi^{2}$ values are reduced by between 7 and 18 for 
the diffuse spectra, but change by only $\Delta \chi^{2} = -1.6$
 to +0.2 in point source 
spectra, and do not
change at all in the ULX spectra. 

A power law continuum model gives a marginally better fit to the diffuse
emission continuum than a bremsstrahlung model in ACIS-S and ACIS-I spectra,
but visual inspection of the best fit models shows that the 
differences between the two best-fit continuum models are very subtle.
It is noteworthy that the slopes of the all of the
best fit power law models to the
diffuse emission ($2.0 \la \Gamma \la 2.7$) are significantly different
from the slope of the resolved point source spectra ($\Gamma = 1.0\pm{0.2}$).
This is further evidence that the bulk of the diffuse hard X-ray emission
does \emph{not} arise in an unresolved population of point sources that
are similar to but fainter than the resolved point source population.
The power law slopes fit for the ACIS-S and ACIS-I diffuse emission
spectra differ at 
greater than 90\% confidence. We will address this issue when we
fit the more elaborate models to the diffuse emission
in \S~\ref{sec:results:spectra:newchandra}.

CCD-resolution X-ray spectra that fit well with a continuum plus weak
line feature are often also well-fit by a continuum plus absorption edge
model. This model is physically plausible in the case of AGN where
there may be very large column densities of absorbing material 
along the line of sight to the point-like continuum source. 
In the case of M82 where the apparent line feature is seen
in the diffuse X-ray emission that covers a region
several hundred parsecs in size an implausible amount of iron would be
required in an absorption edge model. We therefor do not consider this
case further.

\subsubsection{The significance of the iron lines 
in the diffuse emission spectra}

These spectral fits demonstrate the presence of
weak line features in the diffuse emission spectra, and very weak
or negligible line emission in the point source and ULX spectra.
That the ACIS-S and ACIS-I spectra independently yield similar
best-fit  spectral parameters suggests that these results are not a
statistical fluke or calibration oddity. Nevertheless we would like to better 
quantify the true significance of the Fe line detections 
in the diffuse emission, and the lack
of detection in the point source and ULX spectra.

In X-ray astronomy it has been traditional to use an F-test 
to assess the significance of the addition of a weak line component
to a spectral fit, but \citet{protassov02} show that this is
inappropriate. We therefor used Monte-Carlo models to assess the 
significance of the line features in the diffuse and ULX spectra,
as well as the uncertainties in the diffuse emission line fluxes.

\begin{deluxetable*}{lllllllll}
\tablecolumns{9}
\tablewidth{0pc}
\tablecaption{Hard X-ray fluxes and luminosities
        \label{tab:fluxes}}
\tablehead{
\colhead{Dataset} 
    & \multicolumn{2}{l}{Continuum, 2 -- 8 keV)}
    & \multicolumn{2}{l}{6.4 keV Fe K$\alpha$ line}
    & \multicolumn{2}{l}{6.7 keV Fe He$\alpha$ line}
    & \multicolumn{2}{l}{6.9 keV Fe Ly$\alpha$ line} \\
\colhead{\nodata} 
    & \colhead{$f_{X} \times 10^{12}$} & \colhead{$L_{\rm X} \times 10^{-39}$} 
    & \colhead{$f_{X} \times 10^{14}$} & \colhead{$L_{\rm X} \times 10^{-37}$} 
    & \colhead{$f_{X} \times 10^{14}$} & \colhead{$L_{\rm X} \times 10^{-37}$} 
    & \colhead{$f_{X} \times 10^{14}$} & \colhead{$L_{\rm X} \times 10^{-37}$}  \\
\colhead{\nodata} 
    & \colhead{$\ergps \pcmsq$} & \colhead{$\ergps$}  
    & \colhead{$\ergps \pcmsq$} & \colhead{$\ergps$} 
    & \colhead{$\ergps \pcmsq$} & \colhead{$\ergps$}
    & \colhead{$\ergps \pcmsq$} & \colhead{$\ergps$} \\
\colhead{(1)}
    & \colhead{(2)} & \colhead{(3)} 
    & \colhead{(4)} & \colhead{(5)} 
    & \colhead{(6)} & \colhead{(7)}
    & \colhead{(8)} & \colhead{(9)}
}
\startdata
ACIS-S
  & $2.86_{-0.29}^{+0.11}$  &  $4.43_{-0.45}^{+0.17}$ 
  & $5.67_{-3.77}^{+3.77}$  &  $8.78_{-5.85}^{+5.84}$ 
  & $11.00_{-5.02}^{+5.02}$  &  $17.00_{-7.78}^{+7.78}$
  & $<9.24$  &  $<14.3$ \\
33ks ACIS-I
  & $2.76_{-0.12}^{+0.17}$  &  $4.28_{-0.26}^{+0.19}$ 
  & $5.82_{-3.30}^{+3.74}$  &  $9.02_{-5.12}^{+5.80}$ 
  & $7.11_{-4.37}^{+3.83}$  &  $11.0_{-6.78}^{+5.95}$ 
  & $<7.15$  &  $<11.1$\\
Merged ACIS-I
  & $2.81_{-0.12}^{+0.11}$  &  $4.35_{-0.19}^{+0.17}$ 
  & $5.53_{-2.52}^{+3.32}$  &  $8.58_{-3.91}^{+5.15}$ 
  & $8.73_{-3.77}^{+3.36}$  &  $13.50_{-5.85}^{+5.22}$ 
  & $<6.58$  &  $<10.2$\\
XMM-short 
  & \nodata & \nodata
  & $6.77_{-2.30}^{+2.26}$  & $10.50_{-3.57}^{+3.51}$
  & $16.10_{-2.54}^{+2.55}$  & $24.90_{-3.95}^{+3.95}$
  & $7.74_{-2.60}^{+2.59}$  & $12.0_{-4.04}^{+4.02}$ \\
XMM-long
  & \nodata & \nodata 
  & $1.32_{-0.94}^{+0.94}$  & $2.04_{-1.45}^{+1.46}$
  & $8.70_{-1.06}^{+1.10}$  & $13.50_{-1.65}^{+1.71}$
  & $<3.00$  & $<4.66$ \\
ULX high state\tablenotemark{a} 
  & \nodata & \nodata 
  & $5.45_{-2.48}^{+2.45}$  & $8.45_{-3.85}^{+3.80}$
  & $7.37_{-2.76}^{+2.78}$  & $11.40_{-4.28}^{+4.31}$
  & $7.47_{-2.62}^{+2.79}$  & $11.60_{-4.06}^{+4.33}$ \\
\enddata
\tablecomments{Fluxes and luminosities quoted refer to the diffuse emission
alone for the {\it Chandra} ACIS data. The contribution of the unresolved
point sources has been removed from these values. Fluxes derived from the 
{\it XMM-Newton} data are the total line fluxes from the nuclear region,
which are a combination of the point source and diffuse X-ray emission.
No broad-band flux is quoted for the {\it XMM-Newton} data as we do not
attempt to separate the diffuse the diffuse and point source contributions.
The diffuse emission fluxes shown above have been corrected
from those derived from the diffuse spectra to account for the diffuse
flux in the point source and ULX spectra regions (an increase by a factor 
1.32 for the ACIS-S data and 1.20 for both ACIS-I datasets). 
All errors are 68.3\% confidence in one interesting parameter.
Upper limits are reported at 99\% confidence. Note that the E=2 -- 8 keV
energy band continuum luminosities do not include the line emission.
    }
\tablenotetext{a}{This is the excess line flux in the 2001 May 06
observation (XMM-short) above that seen in the 2004 April 21 observation
(XMM-long). This excess is presumed to come from the ULX in the 
high flux state.}
\end{deluxetable*}

We created 40000 realizations of pure-continuum spectra (no lines)
using both the ACIS-S and merged ACIS-I dataset  best-fit
power-law models, and fit them in the same manner as the real
data with power law plus line models. We then assessed how often
the fit-derived line fluxes in the 6.4 keV, 6.7 keV and summed 6.4 and 6.7 keV
lines equaled or exceeded the observed values. We found that we can
reject the null hypothesis (that the lines found in the real data
are merely noise) at high confidence for all lines in both datasets.

For the weaker of the line features seen in the diffuse
emission, the E=6.4 keV line, the null hypothesis can be rejected
at 98.6225\% confidence for the ACIS-S observation alone, and at 
99.9225\% confidence for the merged ACIS-I observation considered on its own. 
The probability that a spurious E=6.4 keV line feature would appear in both 
sets of observations if there were no line emission at
$E=6.4$ and 6.69 keV is essentially negligible.

The null hypothesis can be rejected with even more confidence for
the 6.7 keV line feature, at 99.99985\% confidence for the ACIS-S data
considered alone.
In none of the 40000 simulation of the
merged ACIS-I data did Poissonian noise create a spurious 6.7 keV
line feature with a flux equal to or greater than that observed.

The $E=6.7$ keV Fe line normalization derived from the ACIS-S data
is larger than that from the ACIS-I data by 
 a factor of $\sim 1.6$, marginally larger than
the $68$\% confidence regions. The continuum normalizations
also differ in the same sense but by a factor of $1.2$. 
Accounting for the continuum contribution from unresolved
point sources (discussed below) will reduce the differences in
spectral shape and normalization between the ACIS-S and ACIS-I spectra.

\begin{deluxetable*}{lrllrr}
\tablecolumns{4}
\tablewidth{0pc}
\tablecaption{Best fitting Fe $He\alpha$ line energies \label{tab:line_enrg}}
\tablehead{
\colhead{Dataset} 
    & \colhead{6.4 keV line present?}
    & \multicolumn{2}{c}{Fe He$\alpha$ line}
    & \multicolumn{2}{c}{Fit statistic} \\
\colhead{} & \colhead{}
    & \colhead{Centroid (keV)} & \colhead{Width (keV)}
    & \colhead{$\Delta \chi^{2}$} & \colhead{$\Delta \nu$} \\
\colhead{(1)} & \colhead{(2)}
    & \colhead{(3)} & \colhead{(4)} 
    & \colhead{(5)} & \colhead{(6)} 
}
\startdata
ACIS-S        & yes 
  & $6.66^{+0.10}_{-0.11}$ & (f) & -0.3 & -1 \\
\nodata        & no 
  & $6.66^{+0.10}_{-0.22}$ & (f) & 1.7 & 0 \\
\nodata        & no 
  & $6.61^{+0.14}_{-0.15}$ & $0.15^{+0.19}_{-0.15}$ & 0.2 & -1 \\
Merged ACIS-I &  yes 
  & $6.69\pm{0.17}$ & (f) & -0.1 & -1 \\
\nodata        & no 
  & $6.64^{+0.06}_{-0.22}$ & (f) & 2.6 & 0 \\
\nodata        & no 
  & $6.54^{+0.13}_{-0.15}$ & $0.20^{+0.17}_{-0.20}$ & 0.1 & -1 \\
XMM-long      &  yes 
  & $6.65\pm{0.02}$ & (f) & -6.7 & -1 \\
\nodata        & no 
  & $6.65\pm{0.02}$ & (f) & -5.4 & 0 \\
\nodata        & no 
  & $6.65\pm{0.02}$ & $0.07\pm{0.03}$ & -9.2 & -1\\
XMM-short     &  yes 
  & $6.66\pm{0.03}$ & (f) & -2.3 & -1 \\
\nodata        & no 
  & $6.66^{+0.03}_{-0.04}$ & (f) & 5.6 & 0 \\
\nodata        & no 
  & $6.63^{+0.09}_{-0.08}$ & $0.23^{+0.08}_{-0.07}$ & -2.9 & -1 \\
\enddata
\tablecomments{The change in the fit statistic with respect to the
  best-fit model from Tab~\ref{tab:fits} if the energy and width
  of the $E\sim$ 6.7 keV is fit for, and if the $E=6.4$ keV line
  is present or not. For the {\it Chandra} diffuse emission spectra
  the best-fit model used was the ``PL+PL, three lines'' model.
  For the {\it XMM-Newton} spectra of the entire nuclear region the
  best-fit model was the ``BPL, three lines'' model. Confidence regions 
  correspond to 90\% confidence for one 
  interesting parameter. 
  Column 2: Records whether or not a narrow line at $E=6.4$ keV
    is included in the spectral model.
  Column 3: The energy of the Fe He$\alpha$ Gaussian line centroid (keV).
  Column 4: The Gaussian width of the Fe He$\alpha$ Gaussian line (keV). Note
  that this is $\sigma$, not the FWHM. Models where this width was fixed
  at the default value 0.001 keV are denoted with ``(f)''.
  Column 5: The change in the best-fit $\chi^{2}$ value from the
  default model in Table~\ref{tab:fits}. A negative $\Delta \chi^{2}$ indicates
  a better fit.
  Column 6: The change in the number of degrees of freedom $\nu$
  from the default model.
    }
\end{deluxetable*}

\subsubsection{Is only one of the lines real?}

We also considered the possibility that only one of the lines is genuine,
and that the other is noise. Considering the case where only the 6.4 keV
line was assumed to be genuine, the number of times we found a spurious
6.7 keV line with the observed flux in 10000 simulated spectra
was 1 for the ACIS-S data and 0 for the longer merged ACIS-I observation,
\ie this hypothesis can be rejected at $\ga 99.99$\% confidence for each
of the datasets independently. 

The chance that we could mistakenly find a 
6.4 keV line feature if there were in reality only a 6.7 keV line feature is
\emph{not negligible for the ACIS-S data}: 
this hypothesis can only be rejected at
51\% confidence. In the higher S/N merged ACIS-I observation this hypothesis
can be safely rejected with 99.8\% confidence.

Note that these probabilities assume that the lines are intrinsically
narrow and are at the expected energies of E=6.4 and 6.69 keV. If we allow
the central energy (and optionally the width) of the Gaussian component
represent the Fe He$\alpha$ line to vary the statistical significance of the
6.4 keV line feature is reduced. Table~\ref{tab:line_enrg} shows the best-fit
line energies and $\chi^{2}$ values for models fitting for the
Fe He$\alpha$ line energy compared to the default spectral fits
described above. We also experimented with removing the 6.4 keV line 
so as to assess whether a single line at $E \neq 6.69$ keV provides as good
a fit as two lines at $E=6.4$ and $E=6.69$ keV

For the ACIS diffuse emission spectra fitting for the Fe He$\alpha$ line
energy provides essentially negligible improvement in the fit statistic.
Removing the 6.4 keV line leads to marginally worse fits 
($\Delta \chi^{2} \sim 2$). However a single broader line at 
$E \sim 6.5$ -- 6.6 keV provides as good a fit as two narrow lines
at $E=6.4$ and 6.69 keV in both the ACIS-S and merged ACIS-I diffuse
emission spectra. The best-fitting line width of
$\sigma \sim 0.2\pm{0.2}$ keV is to poorly constrained to be used
as a physical constraint on the validity of the model.

Based on purely statistical arguments we can not distinguish between a model
for the diffuse nuclear emission with two narrow lines at $E=6.4$ and $E=6.69$
keV and a model with a single broader line at $E \sim 6.6$ keV. 
We conclude that
the evidence for the $E=6.4$ keV iron line in the ACIS diffuse emission 
spectra is of marginal statistical significance.

\subsubsection{Iron lines and the point sources}

Including line components in the fits to the summed 
point source and ULX spectra
leads to negligible changes in the best-fit $\chi^{2}$ values. From the
existing fits it is safe to assume that there is no 
significant iron line emission in the spectra of the point sources
(excluding the ULX). In the case of the ULX the apparent continuum
level in the $E=6$ -- 7 keV energy band 
is significantly higher than in the diffuse
or point source spectra. Would such a high continuum make it difficult to see
weak iron line emission in the {\it Chandra} ACIS spectra, or 
even miss line fluxes as high as those
found in the diffuse emission?
For the ULX to be a significant contributor to the total hard X-ray
iron line emission from M82 it would have to produce a
line luminosity comparable to that from the diffuse emission.
We created 10000 simulated
spectra of the ULX, again for both ACIS-S and ACIS-I observations, 
using the best-fit broken power law continuum
model along with iron lines equal in flux to those in the best-fit
diffuse emission spectra, and then evaluated how often we would
have fit line fluxes for the ULX as low we observed. 
For the ACIS-S this would only tend to happen 15.3, 2.5, and 1.0\%
of the time, for the 6.4 keV, 6.7 keV and summed line fluxes respectively.
The equivalent numbers for the ACIS-I dataset are 2.4, 0.7 and 0.1\%.

We conclude that despite the relatively 
high continuum level in the {\it Chandra} ACIS spectra of the ULX
we could detect iron lines 
of comparable flux to that in the diffuse emission spectra with
relative ease. That we did
not in either set of {\it Chandra} observations
indicates that the ULX was not a source of significant 6.4 keV 
or 6.7 keV line emission in September 1999 or June 2002.

\begin{deluxetable*}{llll}
\tablecolumns{4}
\tablewidth{0pc}
\tablecaption{The 6.9 keV line flux and 6.9/6.7 keV line ratio.
        \label{tab:69limits}}
\tablehead{
\colhead{Dataset} 
    & \colhead{6.9 keV line norm}
    & \colhead{6.9/6.7 keV line ratio}
    & \colhead{$kT_{\rm CIE}$} \\
\colhead{(1)}
    & \colhead{(2)} & \colhead{(3)} 
    & \colhead{(4)} 
}
\startdata
ACIS-S        & $<6.27$ & $<0.80$ & $<1.1 \times 10^{8}$ \\
33ks ACIS-I   & $<5.39$ & $<0.97$ & $<1.2 \times 10^{8}$ \\
Merged ACIS-I & $<4.91$ & $<0.72$ & $<1.0 \times 10^{8}$ \\
XMM-long      & $<2.69$ & $<0.33$ & $<7.6 \times 10^{7}$ \\
XMM-short     & $6.9\pm{0.23}$ & $0.46\pm{0.27}$ & $(8.7^{+2.0}_{-2.7}) \times 10^{7}$ \\
ULX high state\tablenotemark{a} 
& $6.7\pm{0.39}$ & $1.03\pm{0.66}$ & $(1.3^{+0.4}_{-0.5}) \times 10^{8}$ \\
\enddata
\tablecomments{Upper limits are reported at 99\% confidence.
  Where two-sided confidence regions 
  are reported they are 68.3\% confidence for one 
  interesting parameter. Column 2: Upper limit or best-fit value for
  the 6.969 keV line normalization, in units of 
  $10^{-6}$ photons s$^{-1}$ cm$^{-2}$. Column 3: Upper limit or best-fit
  value for the ratio of the photon
  flux in the 6.9 keV (Fe Ly$\alpha$) to 6.7 keV (Fe He$\alpha$)  lines,
  Column 4: Upper limit or value for the plasma temperature in Kelvin,
  derived from
  the 6.9/6.7 keV line ratio, assuming collisional ionization equilibrium
  (CIE) and using the APEC plasma code.
    }
\tablenotetext{a}{This is the excess line flux in the 2001 May 06
observation (XMM-short) above that seen in the 2004 April 21 observation
(XMM-long). This excess is presumed to come from the ULX in the 
high flux state.}
\end{deluxetable*}

\subsection{More elaborate fits the Chandra diffuse emission spectra}
\label{sec:results:spectra:newchandra}

It is worthwhile to consider two further elaborations of the previously 
performed spectral fits: accounting for the contribution of unresolved
point sources to the continuum spectral shape, and placing limits on the
line flux of Hydrogen-like Fe at $E=6.97$ keV.

Based on the diffuse fractions described in
 \S~\ref{sec:results:images:fdiff} we know that 
a small but non-negligible fraction of the {\it Chandra} diffuse spectra
is contributed by unresolved point sources (URPS). 
The fraction of the
apparently diffuse count rate in the E=3.0 -- 9.9 keV energy band 
arising in these unresolved sources
is $R_{\rm diff-urps}/R_{\rm diff} = 1 - (f^{\rm T}_{\rm diff}/f_{\rm diff})
\sim 0.23$ for the ACIS-S observation and $\sim 0.13$ for the merged ACIS-I
observation. The longer exposure time of the merged ACIS-I observation 
leads to a significant improvement in point source removal, and reduction
in spectral contamination by unresolved sources. 
The power law slope of the summed spectrum of the resolved  
point sources ($\Gamma \sim 1.0$) is significantly 
flatter that the fitted diffuse emission continua ($\Gamma \sim 2.1$ (ACIS-S)
or 2.7 (ACIS-I)). If the unresolved sources have a similar spectrum to the
resolved point sources then accounting for the URPS contribution should
steepen the fitted continuum spectral shape of the true diffuse emission.

We modeled the contribution of the URPS to the diffuse spectra as a
power law component 
with the same spectral slope as the resolved point sources, and  with
a normalization that produces the expected count rate 
$R_{\rm diff-urps} = (f_{\rm diff} - f^{\rm T}_{\rm diff}) 
\times R_{\rm tot}$ in the 3.0 -- 9.9 keV energy band. 
Given the uncertainties in $f^{\rm T}_{\rm diff}$ and  $R_{\rm tot}$
this normalization is accurate to $\sim 5$\%.
The power
law slope and normalization of the URPS component
are fixed and are not allowed to vary during the spectral fitting.
A Gaussian line component at a fixed energy of $E = 6.969$ keV was
added to represent the Fe Ly$\alpha$ line.  The results
of these fits are given in Table~\ref{tab:fits}.

Both bremsstrahlung and
power law models were used for the continuum shape of the true diffuse
X-ray component, although we found that a power law continuum was
still favored statistically over a bremsstrahlung model. The best fit
normalization of the 6.9 keV line is often zero and in all cases statistically
consistent with zero (although note that the two sided confidence regions 
reported by XSPEC and given in Table~\ref{tab:fits} do not strictly apply
when the region includes zero. Rigorously derived upper limits on the
6.9 keV line flux are
described below).

Best-fit $\chi^{2}$ values are lower (\ie better) than 
the simpler models previously
discussed, but the improvement is not statistically significant. Nevertheless
this exercise is worthwhile in that including the expected
contribution from the URPS leads to a closer match between 
the fitted ACIS-S and ACIS-I diffuse component continuum shape
($\gamma$ or kT).

We used Monte-Carlo simulations (with 5000 realizations each for the ACIS-S,
33ks ACIS-I and merged ACIS-I spectra) to derive upper limits on the 
6.9 keV line flux and 6.9/6.7 line flux ratio. Simulations were used to
convert the limits on the line flux ratio into limits on the plasma
temperature, assuming collisional ionization equilibrium (CIE). These limits
are shown in Table~\ref{tab:69limits}, and the corresponding line luminosities
are given in Table~\ref{tab:fluxes}.

\subsection{XMM spectra}
\label{sec:results:spectra:xmm}

Nuclear region spectra were extracted from both of the XMM-Newton 
observations within a radius of $28\farcs6$ of 
$\alpha=09^{\rm h} 55^{\rm m} 51\fs9, 
\delta=+69\degr 40\arcmin 47\farcs1$ (J2000.0), the same
region as used for the {\it Chandra} spectra. It is impossible to
separate the nuclear diffuse emission from 
the point sources and ULX emission with {XMM-Newton}, so the 
spectra shown in Fig.~\ref{fig:xmmspectra} represent all the emission
from this region. We will not attempt to spectrally-separate the
different components of the diffuse hard X-ray continuum, but are
interested primarily in the Fe line fluxes and energies: are they consistent
with the {\it Chandra} ACIS spectra; and can the superior sensitivity of
the {\it XMM-Newton} spectra be used to detect or place stronger limits
on the $E=6.4$ and $E=6.9$ keV line emission?

For each observation the EPIC PN, MOS1 and MOS2 spectra were simultaneously
fit with the same spectral model, allowing for minor differences between
the PN and MOS calibrations by fitting for a multiplicative term that
is applied to the model normalizations (typically the MOS1 and MOS2 model 
normalizations must be increased by $\sim 5$ -- 8\% [MOS1] and 
$\sim 3$\% [MOS2]). The spectral data are fit only 
in the 3.4 -- 9.0 keV energy range, and the spectra are rebinned to attain
$\ge 10$ counts per bin.

We find that a broken power law provides the best representation of the
continuum spectral shape in both {\it XMM-Newton} datasets. The best fit
spectral parameters are presented in Table~\ref{tab:fits}, the 6.9 keV
line flux and 6.9/6.7 keV line ratios in Table~\ref{tab:69limits}, and the
line luminosities in Table~\ref{tab:fluxes}.

In the longer {\it XMM-Newton} observation (2004 April 21) the $E=6.7$ 
and 6.9 keV iron line fluxes
are very similar to those seen in the Chandra ACIS spectra, although the 
6.4 keV line flux is significantly lower in the {\it XMM-Newton} spectrum.

Fitting for the energy of the Fe He$\alpha$ line in this dataset
results in a significant
improvement in the fit at a line centroid energy of $E= 6.65\pm{0.02}$ keV
(see Table~\ref{tab:line_enrg}). Removing the weak E=6.4 keV Fe K$\alpha$ line
results in slightly worse fit, but as with the ACIS diffuse emission
spectra we find that the evidence for the presence of $E=6.4$ keV emission
is of marginal quality. Once again a broad line (Gaussian 
$\sigma = 0.07\pm{0.03}$ keV, 90\% confidence in one parameter) 
provides a better fit than a narrow Fe He$\alpha$ line.
This, if not due to systematic effects in the detector, corresponds to a
velocity FWHM $\sim 7400\pm{3200} \kmps$ or the thermal line width for
$T \sim 7 \times 10^{10}$ K iron ions. These values are 
large compared to the maximum bulk velocities and temperatures
expected in thermalized SN ejecta of $v \sim 3000 \kmps$ and $T \sim 10^{8}$ 
K \citep{chevclegg}, leading us to doubt whether the broadening of
the $E\sim 6.65$ keV line is real. Unfortunately the lack of a strong
(and intrinsically narrow) 6.4 keV line prevents us from 
assessing the level of systematic effects (for example as done in
\citealt{tanaka00}).

In the {\it XMM-Newton} observation of 2001 May 06 the iron line fluxes
are considerably larger than seen in any of the ACIS spectra or the longer
XMM-Newton observation. Furthermore there is a significant detection of the
6.9 keV line, and the overall hard X-ray continuum is several times
brighter than at the epoch of the XMM-long observation. 
It is most likely that the ULX is
responsible for this additional flux, given previous and more 
extreme hard X-ray variability \citep{ptak99,matsumoto99}.

Assuming that the iron line flux seen in the longer XMM-Newton observation
comes from a non-varying diffuse X-ray component we can calculate the
line flux due to the ULX alone in the high state it was in on 2001 May 06.
For completeness we present these estimated ULX line 
fluxes and a temperature estimate from the 6.9/6.7 keV line ratio 
in Tables~\ref{tab:69limits} \& \ref{tab:fluxes},  even though they do
not pertain to the diffuse emission.

Fitting for the Fe He$\alpha$ line energy in this observation only leads
to a marginal improvement in the fit (Table~\ref{tab:line_enrg}). The 
presence of the 6.4 keV line is strongly favored, if Fe He$\alpha$
is narrow, but as with the other nuclear
spectra we have discussed a single broad line at $E\sim 6.6$ keV
provides as good a fit as two narrow lines at $E=6.4$ keV
and $E=6.69$ keV.

\section{Discussion}
\label{sec:discussion}

The multiple {\it Chandra} and {\it XMM-Newton} observations of M82
provide a relatively self-consistent picture of the
diffuse hard X-ray continuum and iron line emission. 
The emission is brightest within the
spatial region associated with the ongoing starburst and supernova
activity. Emission at $E \sim 6.7$ keV (the Fe XXV He$\alpha$ line
complex) is detected with high statistical significance in all the 
observations we have analyzed. We can only place upper limits on
diffuse $E \sim 6.9$ keV emission (the Fe Ly$\alpha$). The
statistical significance of diffuse $E\sim 6.4$ keV emission is marginal
in both the diffuse spectra obtained with {\it Chandra} and the
total nuclear spectra provide by the {\it XMM-Newton} observations.

\subsection{Iron line emission: previous observations and point sources}
\label{sec:discussion:point_sources}

These results demonstrate that the iron line emission in M82 comes from
a diffuse source and not from compact objects in three of the four 
epochs sampled by {\it Chandra} and {\it XMM-Newton} observations. 
In general the iron line emission from the summed point sources and/or the 
ULX appears to be of negligible
intensity compared to the diffuse iron line emission.

Nevertheless the distinctly higher 
iron line intensity (in particular at $E=6.9$ keV)
seen in the XMM-short observation compared to the relatively consistent
iron line intensities seen the in the other {\it Chandra} and {\it XMM-Newton}
observations indicate that compact objects can contribute at times
to the iron line emission from M82. It is most likely that the additional
iron line emission came from the ULX.

The analysis presented by \citet{strohmayer03}, where they assumed that
all the iron line emission came from the ULX, was based on this XMM-short
observation. With our results we can now separate remove the time-constant
diffuse contribution to the total iron line emission in this observation
(Table~\ref{tab:fluxes}). The 6.4 and 6.7 keV iron line fluxes from ULX
in this high iron-line-emitting state were comparable to the total diffuse
iron line emission. At this epoch the 6.9 keV line emission from the ULX
is of comparable intensity to the 6.7 keV emission.

K-shell iron line emission has previously
been detected in {\it Ginga}, {\it ASCA} 
and {\it BeppoSAX} X-ray spectra of M82, although the uncertainties
in the mean line energy and flux were large.
{\it Ginga} observations of M82 \citep{ohashi90,tsuru92} 
detected emission at  $E\sim 6.7$ keV with an equivalent
width of $170\pm{60}$ eV. \citet{ptak97} reported a marginally
significant
($3\sigma$) detection of a line at $E\sim 6.6$ keV with
an equivalent width of 100 eV based on observations with {\it ASCA}.
Using the same data \citet{tsuru97} concluded that evidence for the
line was not significant, and gave an upper limit at 99\% confidence
on the equivalent width of any line in the energy range
$6.3 \la E$ (keV) $\la 7.1$ of less than 100 eV. Later observations
of M82 with {\it ASCA} found a variable X-ray source (most probably
the ULX) in an extremely luminous state, and in these observations the 
spectra displayed weak
iron line emission with an equivalent width of $121^{+56}_{-61}$ eV
\citep{matsumoto99}. \citet{cappi99} fit for both Fe K$\alpha$ and
Fe He$\alpha$ lines in the {\it BeppoSAX} observations of M82,
of which only the latter was detected (at $E=6.6\pm{0.2}$ keV)
with an equivalent width of $60^{+36}_{-45}$ eV.

These historical values are similar to the
range we find in the two {\it XMM-Newton} observations, where the total
iron line equivalent widths were 109 eV (XMM-short) and 81 eV
(XMM-long), although this may be a coincidence. Indeed, based on
our results we would expect the iron line equivalent width in M82
to vary, as we know that the total hard X-ray continuum is strongly
influenced by the variability associated with point sources, and
also that at some epochs (in one of the four observations we analyze)
one or more point sources also contribute to the iron line emission.

\citet{cappi99} provide Fe He$\alpha$ and Fe K$\alpha$ line photon fluxes and
upper limits in their analysis of the {\it BeppoSAX} observation of M82
which are consistent with the iron line normalizations we find 
(Table~\ref{tab:fits}). The most convincing case for $E=6.4$ keV
Fe K$\alpha$ emission we have is in XMM-short observation. Here the
observed line photon flux was $(6.6\pm{3.7}) \times 10^{-6}$ photons
$\pcmsq \ps$, consistent with the upper limit of $< 1.5 \times 10^{-5}$ photons
$\pcmsq \ps$ in the {\it BeppoSAX} observation. The Fe He$\alpha$ line
flux is not well constrained in the {\it BeppoSAX} data: $(1.6^{+1.1}_{-1.0}) \times 10^{-5}$ photons $\pcmsq \ps$, which is consistent with both the
flux from the diffuse emission alone ($\sim 6$ -- $8 \times 10^{-6}$ photons
$\pcmsq \ps$) or the larger flux of  $(1.5^{+0.4}_{-0.2}) \times 10^{-5}$ 
photons $\pcmsq \ps$ in the XMM-short observation.

Even in {\it Chandra} and {\it XMM-Newton} 
data the iron lines do not immediately
stand out, so it is unsurprising (if disappointing) that the observations
of M82 by previous generations of X-ray telescopes do not place
tight constraints on the historical iron line luminosity of this
galaxy.

\subsection{Expectations for diffuse hard X-ray emission in M82}
\label{sec:discussion:diffuse}

The presence of diffuse hard X-ray emission in the center 
of M82 from a 
$T\sim10^{8}$ K plasma has been expected since the seminal theoretical paper
of \citet{chevclegg}. Various authors have also considered inverse
Compton (IC) scattering of IR photons by relativistic electrons
(responsible for the strong non-thermal radio emission
from the center of M82) as a significant source of
diffuse hard X-ray emission \citep[\eg][]{schaaf89,moran97}.

\subsubsection{6.7 keV iron line emission}
\label{sec:discussion:diffuse:line}

Diffuse thermal X-ray emission is expected in the standard superwind
model \citep{chevclegg,ham90}, and as long as there is little mass-loading
(the mixing of cooler ambient gas into the merged SN ejecta, see \citealt
{suchkov96})
the temperature of the hot plasma in the starburst region will be
high enough for hard X-ray bremsstrahlung emission and 6.7 keV iron line
emission.

Here we will address whether the observed 6.7
keV iron line luminosities imply plasma properties consistent
with a standard \citeauthor{chevclegg} superwind model.
We defer a detailed investigation of what constraints these hard X-ray
observations of M82 can place on starburst superwind models to a 
companion paper.

The highest 6.7 keV line luminosities we can associate with the
diffuse emission 
is $L_{\rm X, 6.7} = 1.70 \times 10^{38} \ergps$ (Table~\ref{tab:fluxes}).
The implied starburst region 
plasma emission integral is $EI = f n_{e} n_{H} V = L_{\rm X, 6.7}/
(Z_{\rm Fe, \star} \Lambda_{6.7}[T_{\star}, Z_{\rm Fe, \odot}])$ where 
the hot plasma of temperature $T_{\star}$,
iron abundance $Z_{\rm Fe, \star}$ (relative to the Solar iron abundance)
and emissivity at Solar abundance $\Lambda_{6.7}[T, Z_{\rm Fe, \odot}]$  
occupies a fraction $f$ of the starburst region volume $V_{\star}$.

Supernova and stellar wind ejecta in a starburst are relatively
enriched in iron with respect to the Solar value, and at the temperatures
concerned it is likely that there is little or no depletion onto
dust grains. To determine  $Z_{\rm Fe, \star}$
we used version 4 of the Starburst99 population synthesis code 
\citep{leitherer99}.
We used a continuous star formation model, with a Salpeter
IMF between stellar masses 1 -- 100 $\Msol$, forming stars of Solar
metallicity, and evaluated the output once the stellar populations are
essentially in a steady state balancing death and replacement 
($\sim 30$ Myr after the onset of star formation). For these
parameters the iron abundance is $Z_{\rm Fe, \star} = 5.2 Z_{Fe, \odot}$ 
on the \citet{anders89} abundance
scale used to derive the X-ray emissivities.
Choosing a different time, or using an instantaneous burst model 
will change this value by less than $\pm{50}$\%.

We assume that all of this iron is in the gas phase. Any iron
initially depleted onto dust grains will be thermally sputtered in
a time scale $n_{e} t \sim 10^{4} \yr \pcc$ for plasma temperatures
$T \ga 10^{7}$ K, and will have reached an ionization state equivalent
to collisional equilibrium in this time \citep{itoh89,smith96,borkowski97}.
We will demonstrate below than $n_{e} t \ga 10^{4} \yr \pcc$ in
the hot plasma at the center of M82, justifying our assumption of
negligible depletion of iron into dust grains.

The 6.7 keV Fe line emissivity peaks at $kT = 5.4$ keV ($T = 6.3 \times 
10^{7}$ K)
at a value of $\Lambda_{6.7} = 9.75 \times 10^{-25} \ergps {\rm cm}^{3}$
per unit Solar iron abundance
(from version 1.3.1 of the APEC plasma code, see \citealt{apec01}).
We will assume that the plasma is at this temperature (which is consistent
with the upper limits given in Table~\ref{tab:69limits}),
but will demonstrate below 
that adopting a different temperature leads to  minor
changes in the derived plasma properties.

The region occupied by the radio SNRs in M82 is $\sim 700$ pc in diameter,
with a projected thickness of $\sim 100$ pc, yielding a total starburst region
volume of $V_{\star} = 1.14 \times 10^{63} {\rm cm}^{3}$.

Adopting the simplifications that $n_{\rm H} \approx 0.83 n_{\rm e}$ 
and that the total number density $n_{\rm tot} = 1.92 n_{\rm e}$, then
the E=6.7 keV line flux in the ACIS-S observation implies that
\begin{equation}
n_{e} \approx 0.19 \, f^{-1/2}
   \, \left( \frac{L_{X, 6.7}}{L_{X, 6.7, \star}} \right)^{1/2}
   \, \left( \frac{Z_{Fe}}{Z_{Fe, \star}} 
   \,  \frac{\Lambda_{6.7}}{\Lambda_{6.7, \star}} 
   \,  \frac{V}{V_{\star}} \right)^{-1/2} \pcc
\end{equation}
where the starred subscript refers to the default values adopted above.
The filling factor of the very hot plasma within the starburst region
is expected to be of order unity (unlike the soft X-ray emitting
plasma in the larger scale wind), so we will assume $f = 1$ for all further
calculations.

Using these adopted values the thermal pressure in the starburst region is
approximately $P_{\star}/k \approx n_{\rm tot} T
\approx 2.3 \times 10^{7}$ K $\pcc$. This is consistent with estimates
of the thermal pressure in warm ionized gas in the starburst
region of $P/k \sim 1$ -- $3 \times 10^{7} \K \pcc$ \citep{ham90,smith06},
although approximately an order of magnitude larger than the values
derived by \citet{forster01}. Preserving the linearity of the
FIR-radio correlation between normal and starburst galaxies
implies that the magnetic field strength in the center of M82 is
$B \sim  200$ -- $400 \mu G$ \citep[See Figs.~1 and 4 of][Thompson, private\
communication]{thompson06}. The resulting magnetic pressure $P_{B}/k 
= B^{2}/8\pi k \sim 1$ -- $4 \times 10^{7} \K \pcc$ would be
in equipartition with the thermal gas pressure. 

The total mass and thermal
energy associated
with the hot plasma within the starburst region is 
$M_{\star} \sim n_{\rm tot} \mu m_{\rm H} f V_{\star} \sim 2.2 
\times 10^{5} \Msol$ and $E_{\rm TH, \star}  \sim 1.5 n_{\rm tot} kT_{\star} 
f V_{\star} \sim 5.5 \times 10^{54} \erg$ (where $\mu \approx 0.61$ 
is the mean mass per particle in the plasma).

The predicted continuum X-ray luminosity in the E=2 -- 8 keV band
associated with this
plasma is $ 2.49 \times 10^{38} \ergps$, approximately a factor
20 less than the observed diffuse continuum. We will address this
issue in greater detail below.

\begin{deluxetable*}{llrrrrr}
\tablecolumns{7}
\tablewidth{0pc}
\tablecaption{Plasma properties inferred from diffuse emission luminosities
        \label{tab:gas_params}}
\tablehead{
\colhead{Parameter} & \colhead{Units}
    & \multicolumn{5}{l}{Dataset} \\
\colhead{\nodata} & \colhead{\nodata} 
    & \colhead{ACIS-S} & \colhead{ACIS-S} & \colhead{Merged ACIS-I}
    & \colhead{ACIS-S} & \colhead{Merged ACIS-I} \\ 
\multicolumn{2}{l}{Input luminosity} 
    & \colhead{$6.7 \keV \, L_{X}$} & \colhead{$6.7 \keV \, L_{X}$}  
    & \colhead{$6.7 \keV \, L_{X}$} & \colhead{Continuum $L_{X}$} 
    & \colhead{Continuum $L_{X}$} \\
\colhead{(1)}
    & \colhead{(2)} & \colhead{(3)} 
    & \colhead{(4)} & \colhead{(5)} 
    & \colhead{(6)} & \colhead{(7)}
}
\startdata
Assumed $kT$     & keV & 5.40 & 3.25 & 5.40 & 3.80 & 2.70 \\
Assumed $T$      & K 
  & $6.3 \times 10^{7}$ & $3.8 \times 10^{7}$ 
  & $6.3 \times 10^{7}$ & $4.4 \times 10^{7}$ 
  & $3.1 \times 10^{7}$ \\
Assumed $Z_{Fe}$ & $Z_{Fe, \odot}$ & 5.2 & 5.2 & 5.2 & 5.2 & 5.2 \\
EI & $cm^{-3}$
  & $3.5 \times 10^{61}$ & $4.5 \times 10^{61}$ 
  & $2.8 \times 10^{61}$ & $8.1 \times 10^{62}$ 
  & $1.1 \times 10^{63}$ \\
$n_{e}$ & $cm^{-3}$
  & 0.19 & 0.22 & 0.17 & 0.93 & 1.08 \\
P/k     & K $cm^{-3}$
  & $2.3 \times 10^{7}$ & $1.6 \times 10^{7}$ 
  & $2.1 \times 10^{7}$ & $7.9 \times 10^{7}$ 
  & $6.5 \times 10^{7}$ \\
$c_{s}$ & $\kmps$
  & 1190 & 922 & 1190 & 997 & 840 \\
$M$ & $M_{\odot}$ 
  & $2.2 \times 10^{5}$ & $2.5 \times 10^{5}$ 
  & $1.9 \times 10^{5}$ & $1.0 \times 10^{6}$ 
  & $1.2 \times 10^{6}$ \\
$E_{\rm TH}$ & $\erg$ 
  & $5.5 \times 10^{54}$ & $3.7 \times 10^{54}$ 
  & $4.9 \times 10^{54}$ & $1.9 \times 10^{55}$ 
  & $1.5 \times 10^{55}$ \\
$E_{\rm KE}/v^{2}_{1000}$ & $\erg$ 
  & $2.2 \times 10^{54}$ & $2.4 \times 10^{54}$ 
  & $1.9 \times 10^{54}$ & $1.0 \times 10^{55}$ 
  & $1.2 \times 10^{55}$ \\
$L_{2-8}$ & $\ergps$ 
  & \tablenotemark{p} $2.49 \times 10^{38}$ 
  & \tablenotemark{p} $2.16 \times 10^{38}$ 
  & \tablenotemark{p} $1.98 \times 10^{38}$ 
  & \tablenotemark{i} $4.43 \times 10^{39}$ 
  & \tablenotemark{i} $4.35 \times 10^{39}$ \\
$L_{6.7}$ & $\ergps$ 
  & \tablenotemark{i} $1.70 \times 10^{38}$ 
  & \tablenotemark{i} $1.70 \times 10^{38}$ 
  & \tablenotemark{i} $1.35 \times 10^{38}$ 
  & \tablenotemark{p} $3.47 \times 10^{39}$ 
  & \tablenotemark{p} $3.28 \times 10^{39}$ \\
$L_{6.9}$ & $\ergps$ 
  & \tablenotemark{p} $4.02 \times 10^{37}$ 
  & \tablenotemark{p} $1.00 \times 10^{37}$ 
  & \tablenotemark{p} $3.20 \times 10^{37}$ 
  & \tablenotemark{p} $3.13 \times 10^{38}$ 
  & \tablenotemark{p} $1.05 \times 10^{38}$ \\
$\tau_{\rm cool}$ & $\yr$ 
  & $3.8 \times 10^{8}$ & $3.0 \times 10^{8}$ 
  & $4.2 \times 10^{8}$ & $7.2 \times 10^{7}$ 
  & $6.3 \times 10^{7}$ \\
$\tau_{\rm flow}$ & $\yr$ 
  & $1.7 \times 10^{5}$ & $2.2 \times 10^{5}$ 
  & $1.7 \times 10^{5}$ & $2.1 \times 10^{5}$ 
  & $2.4 \times 10^{5}$ \\
$\dot M_{\rm cool}$ & $\Msol \pyr$ 
  & $5.7 \times 10^{-4}$ & $8.2 \times 10^{-4}$ 
  & $4.6 \times 10^{-4}$ & $1.5 \times 10^{-2}$ 
  & $1.9 \times 10^{-2}$ \\
$\dot M_{\rm flow}$ & $\Msol \pyr$ 
  & 1.3 & 1.1 & 1.1 & 5.0 & 5.0 \\
$\dot E_{\rm flow}$ & $\ergps$ 
  & $1.0 \times 10^{42}$ & $5.3 \times 10^{41}$ 
  & $8.9 \times 10^{41}$ & $2.9 \times 10^{42}$ 
  & $2.0 \times 10^{42}$ \\
\enddata
\tablecomments{
  Plasma properties for the hot gas in the central region of M82
  derived from either iron line or continuum luminosities, and with
  various assumed plasma temperatures. The derivation of these
  parameters is discussed in \S~\ref{sec:discussion:diffuse}.
  The luminosities quoted are either input (i) from the
  values given in Table~\ref{tab:fluxes}, or predicted (p) based on the plasma
  properties inferred based on the input luminosity and spectral parameters
  assumed. Unit volume filling factor is assumed, and the volume 
  and effective radius of
  the starburst region are $V_{\star} = 1.14 \times 10^{63} {\rm cm}^{3}$
  and $R_{\star} = 210$ pc respectively.
  }
\end{deluxetable*}

Can the starburst in M82 create and maintain such a plasma?
The speed of sound in the
plasma is $c_{\rm s} \approx (\gamma kT / \mu m_{\rm H})^{1/2} \approx
1.2 \times 10^{3}$ km/s. To zeroth order this over-pressured gas
will flow out of the starburst region in a time\footnote{
In practice gas takes longer than the flow timescale 
$\tau_{\rm loss} = R_{\star}/c_{\rm s}$ to flow out of the starburst region. 
In \citeauthor{chevclegg}'s \citeyear{chevclegg} analytical model
gas velocities within the starburst region are a function
of radius, but are always subsonic.
} 
$\tau_{\rm loss} \sim 
R_{\star}/c_{\rm s}$, with a mass flow rate $\Mdot \sim 
M_{\star} c_{\rm s} / R_{\star} \sim 1.3 \Msol \pyr$ (where we have adopted
$R_{\star} \equiv (3 V_{\star} / 4\pi)^{1/3} = 210$ pc). To maintain
the hot plasma the starburst mass injection rate should be greater than
or equal to this value. The associated starburst mechanical energy injection
rate must balance the advective energy loss rate of $\Edot \sim 
E_{\rm TH, \star} c_{\rm s} / R_{\star} \sim 1.0 \times 10^{42} \ergps$.

The IR luminosity of M82 is $L_{\rm IR} = 5.8 \times 10^{10} \Lsol$, 
based on the latest re-calibrated 
12 -- $100\micron$ IRAS fluxes \citep{sanders03}. Using the
formulae in \citet{kennicutt98_review} the star formation
rate associated with this luminosity is $\Mdot_{SFR} = 3.9 \Msol \pyr$
(for a Salpeter IMF between 1 and $100 \Msol$).
For this star formation rate the starburst99 model described above
predicts energy and mass injection rates of 
$\Edot_{\rm SB} = 2.4 \times 10^{42} \epsilon \ergps$ and
 $\Mdot_{\rm SN} = 0.93 \beta \Msol \pyr$, where
$\epsilon$ is the thermalization efficiency and $\beta$ the degree
of mass-loading.

These numbers are not very sensitive to the assumed temperature of the
plasma. In Table~\ref{tab:gas_params} we show the predicted plasma
properties assuming a temperature of $kT = 3.25$ keV, the mean of thermal
bremsstrahlung fits to the hard X-ray continuum. Plasma properties
derived from the 6.7 keV line luminosity in the merged ACIS-I observation
are also very similar to the numbers described above. In all cases 
the predicted bremsstrahlung continuum luminosity 
is significantly less than that of the
observed diffuse hard X-ray continuum.

These estimates demonstrate that the observed $E=6.7$ keV diffuse
iron line emission from M82 is consistent with the starburst superwind
model as long at $\ga 25$\% of the SN mechanical energy is harnessed
by the wind. A more detailed comparison between model predictions and these
observations will appear in a companion paper.

\subsubsection{The mystery of the diffuse continuum}
\label{sec:discussion:diffuse:continuum}

Within 500 pc of the center of M82 we estimate that the total 
diffuse X-ray luminosity is in the range
$L_{\rm X} \approx (4.4\pm{0.2}) \times 10^{39} \ergps$ 
in the $E=2$ -- 8 keV energy band (see Table~\ref{tab:fluxes}).
This luminosity has been corrected to (a) account for the diffuse flux
excluded from the spectrum as because it fell in the regions with detected
point source emission, and (b) account for unresolved point source emission.

This value is higher than the value of $L_{\rm X} = 2.2 \times 10^{39}
\ergps$ quoted by \citet{griffiths2000}, but this difference is to be expected.
The effective area of {\it Chandra} is now known to be 
lower than the values given in the early
calibrations used by \citet{griffiths2000}, and consequently 
the true luminosity is higher. 

It is tempting to interpret the diffuse hard X-ray continuum as
the thermal bremsstrahlung emission associated with the iron line 
emission. Power law models fit the spectra statistically
better than a bremsstrahlung model (Table~\ref{tab:fits}), but inspection
of the fits and their residuals show that the practical differences between
non-thermal and thermal models are small (Fig.~\ref{fig:fitted_spectra}).

This interpretation is problematic for several reasons. The most fundamental
problem with this hypothesis 
is that the equivalent width of the iron line emission 
would then imply a sub-Solar iron abundance if the plasma is
in collisional ionization equilibrium. 
A secondary issue is that
the plasma properties derived assuming the continuum is bremsstrahlung 
emission begin to strain the boundaries of a classic \citeauthor{chevclegg}
starburst model
(even when accounting for emission from unresolved X-ray binaries,
\ie using the BR+PL fits shown in Table~\ref{tab:fits}).

The non-thermal hard X-ray continuum emission due to 
inverse Compton (IC) scattering of IR photons off cosmic 
ray electrons has long been expected to be significant in
starburst regions \citep{schaaf89,moran97,oh01}. In the case of M82
we will show that IC X-ray emission can not be a significant
contributor to the diffuse hard X-ray continuum.

\subsubsection{Collisional equilibrium plasma models for the continuum}
\label{sec:discussion:diffuse:cie}

The equivalent width of the diffuse $E=6.7$ keV emission
 we measure in the central 500
pc of M82 is 0.24 -- 0.32 keV, when corrected for unresolved point 
source emission. For a hot plasma with Solar iron
abundance in collisional ionization equilibrium
the 6.7 keV equivalent width lies in the range 
1.6 -- 0.8 keV for temperatures in the range $7.5 \le \log T \le 7.8$,
which would imply a gas phase iron abundance of $Z_{\rm Fe} \sim 0.15$ -- 
$0.4 Z_{\rm Fe, \odot}$. This is inconsistent with the super-Solar abundance
expected from merged SN ejecta, implying either that the diffuse 
hard continuum is not bremsstrahlung from the iron-line emitting plasma, or
that the plasma has had negligible enrichment beyond that of the 
ambient ISM (again this is under the assumption of collisional 
ionization equilibrium).
As discussed above we do not expect depletion
of iron onto dust to be significant given the densities, temperatures
and flow time scales in the hot plasma.

The metallicity of the stellar population and the warm ISM in M82 is
surprisingly high given its moderate galactic mass: 
Stellar and \ion{H}{2} region O and Ne abundances in M82
are Solar or higher \citep{achtermann95,origlia04,smith06}.
The are very few measurements of iron abundances, the only one we
are aware of is that of \citet{origlia04} who 
derive a stellar iron abundance of $Z_{\rm Fe} = 
0.46^{+0.26}_{-0.17} Z_{\rm Fe, \odot}$ from NIR spectroscopy\footnote{We
note that \citet{origlia04} derived an iron abundance of $Z_{\rm Fe} = 
0.43^{+0.12}_{0.08} Z_{\rm Fe, \odot}$ from the short {\it XMM-Newton}
observation, for which we have shown the ULX contributes significantly to
the continuum and iron line emission. \citeauthor{origlia04} 
attempted to correct for point source emission using the
{\it Chandra} ACIS-I observations, 
but given the difference in ULX states between the
different observations this would not have worked. 
It is purely a coincidence that their iron abundance estimate is
similar to that we estimate for the diffuse emission alone.}.
This is consistent with the largest values derived from the iron
line equivalent width, so we will entertain the possibility that
the diffuse continuum emission arises in a hot plasma.

As can be seen
in Table~\ref{tab:gas_params} 
the best fit bremsstrahlung temperatures (ACIS-S: $kT=3.8$ keV;
merged ACIS-I: $kT=2.7$ keV, see Table~\ref{tab:fits}) 
and luminosities (Table~\ref{tab:fluxes}) would indicate plasma volume emission
measures well over one order of magnitude greater than
those derived from the iron lines alone. Other derived plasma properties
(\eg mass, energy)
are larger than those derived from the iron lines by a smaller factor, but
still the energy flow rates equal or exceed the levels that can be 
supplied by the starburst. The implied thermal pressures are now higher
than independent estimates of the pressure in the starburst region. These
discrepancies alone suggest 
that the diffuse hard continuum is not predominantly thermal
bremsstrahlung.

Returning to the issue of the metal abundance it would be difficult for
SNe to heat an ambient plasma to $kT \ga 3$ keV 
without mixing in metal enriched ejecta. To
obtain a final iron abundance of $Z_{\rm Fe}\sim 
0.4 Z_{\rm Fe, \odot}$ (\ie very close to the ambient value) 
and hide any signature of enrichment by
 $Z_{\rm Fe} \sim 5 Z_{\rm Fe, \odot}$ ejecta
would require a large mass ratio of ambient gas to ejecta, 
$M_{\rm amb}/M_{\rm ej} \ga 10$. But if such heavy mass loading
occurred the maximum gas temperature would be $kT \la 1$ keV,
inconsistent with the spectral fits to the continuum and also
inconsistent with the detection of any E=6.7 keV line emission.

We therefor conclude that it is unlikely that both the iron line
emission and the diffuse hard continuum arise in a hot plasma
in collisional ionization equilibrium.

\subsubsection{Non-equilibrium plasma models for the continuum}
\label{sec:discussion:diffuse:nie}

If the iron ion populations are significantly far from collisional
ionization equilibrium 
values it might be possible to reconcile the iron line emission with
the diffuse continuum.
The best-fit Fe He$\alpha$ line energies in Table~\ref{tab:line_enrg},
although for the most part statistically consistent with the expected
value of $E=6.69$ keV,
lie in the range $E\sim 6.54$ to $E=6.66$ keV which is suggestive
of a non-equilibrium plasma.
Comparing the observed iron line fluxes with
those predicted from the continuum we see that
a Fe XXV ion population fraction
about 1.3 dex below that expected for a CIE plasma would be
required.

Electron and ion kinetic temperatures are far from equilibrium in 
supernova remnants with high velocity shocks 
\citep[see \eg][]{vink03,raymond05}. In a collision-less
shock each particle species $s$ thermalizes to a temperature 
$T_{s} = 2.4 \times 10^{5} \, (m_{s}/m_{p}) (v/100 \kmps)^{2}$,
so we would expect iron to initially have a temperature significantly
higher than the electron or proton temperature, before relaxing
into equilibrium via Coulomb collisions or on
a faster timescale by plasma instabilities.

For a electron number density of $n_{e} = 0.93 \pcc$ and
temperature $T = 4.4\times 10 ^{7}$ K (values appropriate if the
continuum is bremsstrahlung, see Table~\ref{tab:gas_params}),
the proton-proton Coulomb
relaxation timescale is $\tau_{pp} \sim 3.2 \times 10^{3}$ yr
\citep{spitzer62}. The electron-electron relaxation timescale is shorter
by a factor $(m_p/m_e)^{1/2} \sim 43$. Assuming that the iron temperature
was initially $m_{Fe}/m_{p}$ times higher than that of the protons
then the Fe-proton relaxation timescale
is still short: $\tau_{Fe-p} \sim 400$ years. These time scales are short
compared to the $10^{5}$ -- $10^{6}$ year time scales significant within
the starburst region, suggesting that the kinetic temperatures 
of the ions and electrons should be very similar $T_{Fe} \approx 
T_{p} \approx T_{e}$.

A given ion will approach CIE on a timescale
\begin{equation}
\tau_{eq} \sim [n_{e} \, (C_{i} + \alpha_{i})]^{-1}
\end{equation}
where $C_{i}$ is the collisional ionization rate coefficient
from charge state $i$ to state $i+1$, and $\alpha_{i}$ is the
total recombination rate coefficient out of state $i$ into state
$i-1$ \citep{liedahl99}. We calculated these rate coefficients
for Fe XXV using the atomic data and methods given in
in \citet{ar85} and \citet{ar92}, although
we used a more accurate approximation for the first exponential
integral (Equations 5.1.53 and 5.1.56 in \citealt{abramowitz+stegun})
than the one given in \citet{ar85}. For $T = 4.4 \times 10^{7}$ K we
find $n_{e} \tau_{eq} = 1.7\times 10^{4} \yr \pcc$.
These rate coefficients vary slowly
with temperature, so hot iron
in M82's starburst region should be close to CIE as 
long at $n_{e} \ga 0.1 \pcc$.

All of these estimates suggest that non-ionization equilibrium
conditions and order of magnitude changes in the Fe XXV ion fraction
are unlikely in the iron-line-emitting plasma in M82. We can not
explain why the best-fit
Fe He$\alpha$ line energies are systematically lower than the collisional
equilibrium value of $6.69$ keV unless it is a statistical fluke, 
as systematics would have to effect both the {\it Chandra} and 
{\it XMM-Newton} data. 
We are left
with the conclusion that the diffuse hard X-ray continuum is not
thermal bremsstrahlung associated with the iron line emission.

\subsubsection{The continuum as inverse Compton emission}
\label{sec:discussion:diffuse:ic}

The simplest non-thermal process capable of generating diffuse hard
X-ray emission in a starburst is inverse Compton scattering of IR
photons by cosmic ray electrons.

However, we do not believe that the dominant process creating the
observed diffuse continuum in the center of M82 can be IC X-ray emission.
\citet{klein88} show that the 
radio spectrum of M82 has a spectral index 
of $\alpha \sim -0.3$ at frequencies
below 1 GHz, and -0.7 above 1 GHz. Assuming that mean wavelength
of the IR photons is $80\micron$ then 
IC X-rays in the 2 -- 8  keV energy band are created by
scattering off cosmic ray electrons with $\gamma$ in the range 
$360 \le \gamma \le 720$. Using the 1.49 GHz flux density 
of 7.66 Jy and size of $36\arcsec \times 18 \arcsec$ \citep{condon90},
 the resulting minimum energy magnetic
field strength is $B_{\rm min} \sim 55 \mu$G (assuming the ratio
of energy in protons to electrons is 100). This implies that
the electron population responsible for the
synchrotron radiation at $\nu \sim 20$ -- 80 MHz is the
source of any 2 -- 8 keV IC X-ray emission.
The resulting IC X-ray spectrum should have a photon index $\Gamma 
= 1 - \alpha \sim 1.3$. The {\it Chandra} observations constrain 
the X-ray continuum photon index 
$\Gamma = 2.5$ -- 3.0 ($\pm0.5$ at 90\% confidence), which is inconsistent 
with the $\Gamma = 1.3$ slope that would be the signature of IC X-ray 
emission.

We can also estimate the IC X-ray luminosity in the hard X-ray band
given that the ratio of the IC to radio synchrotron luminosities is
is equal to the ratio of photon to magnetic field density,
$L_{X, IC}/L_{\rm SYNC} = U_{PH}/U_B$. To first order the IR photon energy
density $U_{PH} \sim L_{IR} / (4\pi c \, R_{\star}^{2}) = 1.4 \times 10^{-9} 
\erg \pcc$. The (minimum) energy density in the magnetic field is considerably
smaller, $U_{B} = B^{2}/8\pi = 1.2 \times 10^{-10} \erg \pcc$.
There is some uncertainty in what value to adopt for the radio synchrotron
luminosity, given that the flatter spectral slope at frequencies
below $1$ GHz suggests absorption. Nevertheless the emission
is not optically thick, as the observed slope of 
$\alpha \approx -0.3$ differs from the $\alpha = +0.5$ of an optically
thick source. With no correction for absorption the
radio spectral energy distribution given by \citet{klein88} 
implies $L_{\rm SYNC} = 2.0 \times 10^{37}
\ergps$ in the 20 -- 80 MHz band. If we extrapolate from fluxes
above 1 GHz assuming the intrinsic spectral slope is $\alpha = -0.7$
then we obtain  $L^{\prime}_{\rm SYNC} = 8.8 \times 10^{37}
\ergps$. The predicted IC X-ray luminosities in the
E=2 -- 8 keV energy band are $L_{IC} = 2.3 \times 10^{38} \ergps$
and $L_{IC}^{\prime} = 1.0 \times 10^{39} \ergps$. These values
are 4 to 20 times smaller than the observed diffuse X-ray continuum,
 and cement our conclusion that IC X-ray emission is not
responsible for the diffuse hard X-ray continuum in M82.


\citet{thompson06} argue that the linearity of the FIR-radio correlation
in galaxies indicates that the minimum energy magnetic field strength
underestimates the true magnetic field strength in starbursts. If this
is true then our estimates of the expected IC X-ray luminosity are
too high. 

The physical origin of the observed X-ray continuum
in M82 is not clear to us at the present time.

\subsection{Similarities to the center of the Milky Way?}

Another peculiarity of the diffuse hard X-ray emission in M82
is that emission from nearly neutral iron (the $E\sim6.4$ keV line emission)
is not predicted in the standard models for X-ray emission
from a starburst region. If the emission at 6.4 keV in M82 is real 
and not a statistical artifact, then the X-ray emission from
center of M82 does bear some
resemblance to the diffuse hard X-ray emission seen in the 
Milky Way Galactic Ridge
and Galactic Center.

The Galactic Ridge X-ray Emission 
\citep[GRXE, see \eg][]{warwick80,yamauchi93} is
an apparently diffuse source of both hard X-ray continuum radiation and
Fe line emission with a vertical scale height of $\sim 100$ pc
is found in the inner $|l| \la 40\degr$ 
of the Milky Way. 
This emission has not been resolved into individual point sources
in deep {\it Chandra} observations and hence might be truly
diffuse in origin (see \eg
\citealt{ebisawa01,ebisawa05}), although \citet{revnivtsev05}
present a strong argument in favor of unresolved point sources 
as the origin of the GRXE.

The GRXE has a 
total $E=2$ -- 10 keV energy band luminosity of $L_{\rm X} =$ 
(1--2) $\times 10^{38} \ergps$ \citep{koyama86,tanuma99} and a total iron line
luminosity of $\sim 9 \times 10^{36} \ergps$ \citep{yamauchi93},
$\sim$ 30 \% of which comes from within 
$\sim 2 \degr$ ($\sim 560$ pc) of the Galactic Center 
\citep{koyama89,yamauchi90}. 
These values are $\sim 5$\% of both the 
diffuse hard X-ray continuum and iron line luminosities we derive
for M82.

Line emission from $E \sim 6.4$ keV (Fe K$\alpha$), 
$E \sim 6.7$ keV (Fe He$\alpha$) and $E \sim 6.9$ keV (Fe Ly$\alpha$) 
line is are often seen, although the relative line intensities vary with
position \citep{koyama96,yusefzadeh02,park04,muno04}. 
In this respect the GRXE and the emission from the 
Galactic Center (GC) differ from the nuclear X-ray
emission from M82, where the $E=6.9$ keV
line is not clearly detected.

The spectrum of the 
hard X-ray continuum near the GC can be fit by either $kT\sim 8$ keV
thermal models or $\Gamma \sim 1.9$ power law models
\citep{muno04}. This is a harder spectrum than the M82
hard diffuse continuum ($kT \sim 3$ -- 4 keV, $\Gamma = 2.5$ -- 3.0).

Models which attempt to explain the GR and GC continuum
emission as thermal bremsstrahlung from a $kT \sim 10$ keV hot plasma
are not currently favored \citep{tanaka00}.
A plasma this hot is not gravitationally bound by the Galaxy, and given the
large Galactic volume the GRXE appears to occupy any thermal
model must to have a large mechanical energy input to balance the advective
losses. \citet{dogiel02} estimate, by assuming the continuum is 
thermal bremsstrahlung emission, that the mechanical power 
required in a thermal bremsstrahlung model for the GRXE is 
$\Edot \sim 10^{43} \ergps$. This exceeds the mechanical energy input from
all supernovae in the Galaxy even assuming high thermalization
efficiency $\epsilon$.

It is also known that the intensity of $E \sim 6.4$ keV line emission 
is highest in the vicinity of molecular clouds
\citep{koyama96,murakami00,murakami01,yusefzadeh02,park04}. 
The X-ray emission in these regions is well fit by a reflection
and fluorescent iron line emission model, although none of the
observed X-ray sources is currently bright enough to be the original
source of the reflected X-ray emission \citep{koyama96,murakami00}.
Diffuse hard X-ray emission from the Sagittarius B2 cloud is well
explained by irradiation from the direction of the GC by a transient source 
of X-ray luminosity $L_{\rm X, 2-10} \sim 3 \times 10^{39} \ergps$ ($E=2$ 
-- 10 keV) \citep{murakami01}.

Other models for the diffuse iron line emission from the GC
include excitation of neutral iron by impact ionization by
low energy Cosmic Ray electrons \citep{valinia00,yusefzadeh02} or charge
exchange processes \citep{wargelin05}. 
Another alternative is non-equilibrium emission excited in a 
$kT\sim 0.6$ keV plasma by quasi-thermal electrons \citep{masai02}.
The non-thermal mechanisms of \citet{masai02}
and \citet{wargelin05} also produce $E=6.7$ and 6.9 keV
iron emission, and so might explain all the observed iron line
emission.
These exotic models appear plausible on energetic grounds, but more 
a set of rigorous predictions and observations will be required to test
these hypotheses.

So there are puzzling similarities and differences between
the diffuse iron line and continuum emission in M82 and the larger
scale GRXE in the Milky Way.
High spectral resolution observations with the proposed
calorimeters on Constellation-X are the only robust way to
exclude the possibility that the iron line emission from M82
comes from an exotic non-thermal mechanism and
is not wind-related thermal emission. It is clear that  
analogies between the normal spiral Milky Way and the 
extreme starburst superwind galaxy M82 must cease to be appropriate
at some point.
The existence of a large scale superwind in M82 directly
requires that there should be a very hot plasma
within the starburst region, unless our understanding
of starburst-driven winds is deeply flawed.

Deeper X-ray observations
with the current generation of X-ray telescopes would be
beneficial, in particular for better constraining the
existence and fluxes of the weaker 6.4 and 6.9 keV lines.
We speculate that the diffuse Fe K$\alpha$ emission from M82, if
indeed present, might arise in a scaled-up version of the X-ray
reflection nebulae seen near Galactic Center molecular clouds 
\citep[\eg][]{koyama96,murakami00}. The most-promising radiation
source for such a model would be 
the M82 ULX, which exhibits flares in X-ray luminosity up to
$L_{\rm X, 2-10} \sim 6 \times 10^{40} \ergps$, assuming
isotropic emission \citep{ptak99,matsumoto99}.

With regard to the extended 6.7 keV iron line emission from
M82 (but not the GC) we conclude that the most plausible explanation for it
remains that it is thermal emission associated with
wind fluid of merged SN ejecta in the starburst region.

\section{Conclusions}
\label{sec:conclusions}

We obtained new {\it Chandra} ACIS-S observations of M82, along with 
reanalyzing two earlier ACIS-I observations and
two {\it XMM-Newton} observations from the public archive.

Our aim was to re-investigate the diffuse hard
X-ray emission discovered by \citet{griffiths2000} using
the best current calibration and CTI correction methods, 
exploiting both the high spatial resolution of
{\it Chandra} and the high sensitivity of the {\it XMM-Newton} EPIC
detectors.

With Chandra we can
spatially separate the diffuse hard X-ray emission from resolved
point source emission, and by using the point source luminosity function 
we also correct for the contribution of lower luminosity unresolved point
sources to the apparently diffuse emission. The XMM-Newton observations
lack the spatial resolution necessary for removing the point sources
within the nuclear regions of M82, but provide high S/N spectra 
of the summed diffuse and point source emission.

Approximately 20 X-ray point sources down to a limiting luminosity of 
$L_{\rm X} \sim 4 \times 10^{36} \ergps$ 
(0.3--8.0 keV energy band) are detected within the central
region of M82. We estimate that the
ACIS-S observation resolves $\sim 90$\% of all point source X-ray emission
from M82 ($\sim 93$\% for the combined ACIS-I exposures). We remove
and interpolate over the resolved point sources to obtain estimates
of the flux from truly diffuse soft and hard X-ray emission in M82.
Correcting for unresolved point sources we estimate that 20 -- 30\%
of the hard X-ray emission ($3 \le E$ (keV) $\le 10$) is truly
diffuse, with a $E=2$ -- 8 keV luminosity of $L_{\rm X, 2-8 keV} \sim
4.4 \times 10^{39} \ergps$. The diffuse continuum is slightly better
fit with power law models (best fit photon index $\Gamma = 2.5$ -- 3.0)
than with thermal bremsstrahlung emission (best fit $kT \sim 3$ -- 4 keV).

The diffuse hard X-ray emission is brightest within the starburst
region, with a major axis extent well matched to the extent of 
the region in which the compact radio sources (young SNRs) 
have been observed. The diffuse hard
X-ray emission is relatively smoothly distributed within this region,
suggestive of a larger volume filling factor than that of the
soft X-ray emitting plasma. Strong, spatially-varying absorption
is visible in the soft X-ray band within the plane of the galaxy,
and this is likely to complicate X-ray-derived abundance estimates.



The diffuse emission spectra from both ACIS-I and ACIS-S observations
obtained within $r = 500$ pc of the nucleus of M82 definitely requires
the presence of a $E \sim 6.7$ keV line component (emission from
highly ionized iron), and possibly the presence of $E=6.4$ keV
Fe K$\alpha$ emission (from neutral or nearly neutral iron). 
We used Monte Carlo simulations to assess 
the significance of these detections. 
The total nuclear region iron line fluxes in the 2004 April 21 
XMM-Newton observation
are consistent with those of the Chandra-derived diffuse component.
We place upper
limits on the diffuse $E\sim 6.9$ keV Fe Ly$\alpha$ line,
which constrain the plasma temperature $T \la 8 \times 10^{7}$ K.

No statistically significant
Fe emission is found in the summed X-ray spectra of the point-like X-ray
sources or the ULX in the two epochs of {\it Chandra} observation.
The total nuclear region iron line fluxes in the 2001 May 6 
{\it XMM-Newton} observation (previously studied in \citealt{strohmayer03})
are significantly higher than the
other {\it Chandra} and {\it XMM-Newton} observations.
We attribute the excess iron line emission in this observation,
where $E=6.4$ keV and $E=6.9$ keV emission lines are also seen, 
to the Ultra-Luminous 
X-ray source in its high state. The iron line luminosity of the
ULX in this flare state was approximately equal to the diffuse
iron line luminosity of the starburst region. 
However, at most epochs of observation (when the ULX was 
not in a high luminosity state)
the iron K-shell luminosity of M82 is
dominated by the diffuse component, contradicting the conclusions
of \citealt{strohmayer03}.

The luminosity associated with the diffuse
iron lines is $L_{\rm X, 6.7 keV} =$ (1.1 -- 1.7) $\times 10^{38} \ergps$
and $L_{\rm X, 6.4 keV} \sim$ (2 -- 9) $\times 10^{37} \ergps$ 
(note that the 6.4 keV line emission is of marginal statistical
significance). We consider the similarities and differences between the
diffuse iron line emission in M82 and the Milky Way Galactic Ridge
X-ray Emission, and conclude that a thermal model is the most
plausible explanation for the 6.7 keV emission in M82.
If we assume that the hot plasma generating the 6.7 keV iron line
is enriched with iron to the degree expected for SN ejecta in
a starburst ($Z_{\rm Fe} \sim 5 \, Z_{\rm Fe, \odot}$) then
the luminosity of the 6.7 keV iron line implies
a gas pressure of  $P/k \sim 2 \times 10^{7} \K \pcc$, and
requires mass and energy injection rates of $\Mdot 
\sim 1.3 \Msol \pyr$ and $\Edot \sim 1.0 \times 10^{42} \ergps$
to maintain. These values are consistent with independent
estimates for the starburst in M82, supporting our hypothesis
that the diffuse 6.7 keV iron line emission arises in the
long-expected but previously unobserved wind fluid that drives
the larger scale superwind in this galaxy.

The nature and origin of the diffuse hard continuum is harder to ascertain.
It is difficult to convincingly interpret the continuum
as the thermal bremsstrahlung associated with the 6.7 keV iron line
emission. For a plasma in collisional ionization equilibrium this would require
a low gas-phase iron abundance 
$Z_{\rm Fe} \la 0.4 Z_{\rm Fe, \odot}$, much lower than the super-Solar
abundance expected for supernova ejecta. The pressure and energy flow
rates derived from the continuum also strain or exceed the
standard starburst-driven wind model. Mass loading could not
produce such a low iron abundance without
cooling the gas to such a degree that no 6.7 keV line emission would
be produced.
We explore the possibility of non-equilibrium ionization of the iron, 
but find that a variety of methods show that collisional
equilibrium is expected in the iron-line-emitting plasma.
Non-equilibrium conditions
 can not reconcile the continuum and iron line emission.
The steep slope of the continuum (which is quite different from the
continuum slope of the point sources, $\Gamma \sim 1$) 
is strong evidence against Inverse Compton
X-ray emission being the dominant source of the diffuse continuum.

The logical next steps are to see what constraints these observations
place on theoretical models for superwinds (in particular the
efficiency of supernova heating the degree of mass loading), and
test these models against observed 6.7 keV iron line fluxes for
other starburst galaxies.

\acknowledgements DKS acknowledges funding through grant G02-3108X 
awarded by the Smithsonian Astrophysical Observatory on behalf of NASA,
and grant NNG05GF62G of the NASA Astrophysics Theory Program. 
We thank Uli Klein, Julian Krolik, Nancy Levenson, 
Andrew Ptak, Leisa Townsley, 
Todd Thompson, Bob Warwick and Farhad Yusef-Zadeh
for helpful comments and advice. We would also like to thank
the anonymous referee for their insightful and helpful report.


\bibliography{dks_refs}

\end{document}